\documentclass[twocolumn,onecolappendix]{aastex631}

\UseRawInputEncoding

\usepackage{amsmath}
\usepackage{soul}

\shorttitle{HOPS 358 Sub-mm Variability}
\shortauthors{Sheehan et al.}

\begin{document}

\title{Sub-Millimeter Variability in the Envelope \& Warped Protostellar Disk of the Class 0 Protostar HOPS 358}

\author[0000-0002-9209-8708]{Patrick D. Sheehan}
\affiliation{National Radio Astronomy Observatory, 520 Edgemont Rd., Charlottesville, VA 22903, USA}

\author[0000-0002-6773-459X]{Doug Johnstone}
\affiliation{NRC Herzberg Astronomy and Astrophysics, 5071 West Saanich Rd, Victoria, BC, V9E 2E7, Canada}
\affiliation{Department of Physics and Astronomy, University of Victoria, Victoria, BC, V8P 5C2, Canada}

\author[0000-0003-1894-1880]{Carlos Contreras Pe\~{n}a}
\affiliation{Department of Physics and Astronomy, Seoul National University, 1 Gwanak-ro, Gwanak-gu, Seoul 08826, Korea}
\affiliation{Research Institute of Basic Sciences, Seoul National University, Seoul 08826, Republic of Korea}

\author[0000-0001-6324-8482]{Seonjae Lee}
\affil{Department of Physics and Astronomy, Seoul National University, 1 Gwanak-ro, Gwanak-gu, Seoul 08826, Korea}

\author[0000-0002-7154-6065]{Gregory Herczeg}
\affiliation{Kavli Institute for Astronomy and Astrophysics, Peking University, Yiheyuan Lu 5, Haidian Qu, 100871 Beijing, Peoples Republic of China}
\affiliation{Department of Astronomy, Peking University, Yiheyuan 5, Haidian Qu, 100871 Beijing, China}
 
\author[0000-0003-3119-2087]{Jeong-Eun Lee}
\affiliation{Department of Physics and Astronomy, Seoul National University, 1 Gwanak-ro, Gwanak-gu, Seoul 08826, Korea}
\affiliation{SNU Astronomy Research Center, Seoul National University, 1 Gwanak-ro, Gwanak-gu, Seoul 08826, Korea}

\author[0000-0002-6956-0730]{Steve Mairs}
\affiliation{NRC Herzberg Astronomy and Astrophysics, 5071 West Saanich Rd, Victoria, BC, V9E 2E7, Canada}
\affiliation{East Asian Observatory, 660 N. A`oh\={o}k\={u} Place,
Hilo, Hawai`i, 96720, USA}

\author[0000-0002-6195-0152]{John J. Tobin}
\affiliation{National Radio Astronomy Observatory, 520 Edgemont Rd., Charlottesville, VA 22903}

\author{Hyeong-Sik Yun}
\affiliation{Korea Astronomy and Space Science Institute, 776 Daedeok-daero, Yuseong, Daejeon 34055, Korea}

\author{The JCMT Transient Team}

\begin{abstract}
    The JCMT Transient Survey recently discovered that the Class 0 protostar HOPS 358 decreased in 350 GHz continuum brightness by $\sim25$\% over the course of four years before brightening again for the next four. The JCMT lightcurve can be fit by a long 
    {timescale dip lasting roughly eight years.}
    A 
    {shorter timescale periodicity} is also apparent with a period of 1.75 years and a 
    {small} 3\% amplitude. NEOWise monitoring reveals that the mid-IR wavelength brightness of HOPS 358 follows a similar long-term pattern in time. Here, we present a study of nine epochs of ALMA observations of HOPS 358 taken over the course of the decline and subsequent rise in brightness seen with the JCMT to test whether the variation seen on $\sim15"$ scales, covering both disk and envelope, is also observed on smaller, $<1"$ scales that primarily probe HOPS 358's protostellar disk. We detect both HOPS 358 and its southern companion, HOPS 358B, in our ALMA observations, and find that at least one of the two is varying. Assuming that HOPS 358 is the variable, the light curve has the same shape as that found by the JCMT.
    Additionally, our high resolution ALMA imaging of HOPS 358 reveals that the disk is warped, with a $16^{\circ}$ warp at a disk radius of 35 au, about halfway through the extent of the disk. The physical origin of the warp along with how it relates to the variability seen towards HOPS 358, however, remain unclear.
\end{abstract}


\section{Introduction}

A long standing problem in star formation is how stars gain the majority of their mass during the main accretion phase, while they are still deeply embedded in their natal envelopes \citep[$\lesssim0.5$ Myr; e.g.][]{Evans2009TheLifetimes,Dunham2014OnDiscs,Kristensen2018ProtostellarEstimates}. Early estimates of protostellar accretion rates, calculated by assuming that protostellar luminosities come from the steady release of gravitational energy as mass accretes onto the protostar, were much too small to explain the growth to stellar masses over the assumed timescale of $\sim0.1$ Myr for stars to form \citep[e.g.][]{Kenyon1990AnCloud,Dunham2010EvolutionaryObservations}. 
Upward revisions to the timescale for stars to form that have been suggested \citep[e.g.][though notably disputed by, e.g. \citet{Kristensen2018ProtostellarEstimates}]{Evans2009TheLifetimes} along with improvements to sensitivity and sample size have tended to significantly alleviate this original ``luminosity problem" \citep[e.g.][]{Offner2011THEFUNCTION,Fischer2023AccretionAssembly}. {Instead, today, more work is focused on explaining the} ``protostellar luminosity spread"{, which is that the observed luminosities of protostars span 3 -- 4 orders of magnitude. Simple models featuring constant accretion rates have difficulties explaining this spread. Models with mass- and time-variable accretion can explain the spread, but whether those variations are steady, variable, or some combination remains the subject of much study and debate \citep[for further details, see][]{Fischer2023AccretionAssembly}.} Thus, there is still much to learn about accretion of mass onto protostars, including how variable it is across time.

Importantly, once a protostar begins to build a protoplanetary disk, the mass accretion rate onto the protostar is determined by the effective viscosity throughout the disk \citep[see review by][]{Hartmann2016AccretionStars}. There is no inherent reason to expect this accretion to be smooth or stable across either radius or time. Measurement of variations in the accretion rate onto the protostar and the commensurable appearance of the disk, therefore, provide vital clues to the physical conditions regulating mass transport within the disk along with protostellar mass assembly \citep{Lee2020SpiralAccretion}.

Direct evidence for variability from deeply embedded protostars has been difficult to find. Although outbursts in a number of young sources have been observed \citep[e.g. FUori and EX\,Lup-type events; e.g.][]{Wachmann1939Spektral-DurchmusterungMilchstrassenfeldern.,Herbig1977EruptiveEvolution,Aspin2010TheLupi}, such outbursts have primarily been seen in sources nearing the end of the accretion phase \citep[e.g.][]{Fischer2023AccretionAssembly}. At earlier times, when most of a star's mass is thought to be accreted, the high extinction from the envelope makes it difficult to directly observe the protostar in the optical or near-infrared. Consequently, only a few strong outbursts from very young sources have been identified \citep[e.g.][]{Fischer2012MULTIWAVELENGTHREGIME,Safron2015HOPSORION,Hunter2017AnContinuum,Zakri2022TheOrion}. More recently, mid-infrared monitoring of Class 0/I sources by {space-based telescopes such as {\it Spitzer} and} {\it WISE} has allowed for cataloging long-term variability \citep{Park2021QuantifyingExplorer} and searches for additional burst events \citep{Fischer2019ConstrainingClouds,Zakri2022TheOrion, 2023Contreras_b,Wang2023AJ064722.95+031644.6, Tran2024WTP10aaauow:Data}.

The youngest, most deeply embedded, Class 0 sources are bright in the sub-millimeter (submm). \citet{Johnstone2013ContinuumStructure} showed that the significant increase in accretion luminosity resulting from an outburst, as well as the subsequent dimming, will lead to measurable changes in the temperature of the dust residing within the protostellar envelope, and that this temperature change should be observable as a change in observed continuum brightness from the far-infrared through submm. To build on this result, the JCMT Transient Survey \citep{Herczeg2017HowRegions,Mairs2024The2.0} has been monitoring with monthly cadence eight star-forming regions over the past eight years to search for variability in the sub-millimeter dust continuum emission from protostellar envelopes. Significant variability has already been uncovered in a handful of sources, including both a quasi-periodic light curve \citep[Class I V371 Ser (EC\,53);][]{Yoo2017TheMain,Lee2020YoungDisk} and a months-long burst \citep[Class 0 HOPS-373;][]{Yoon2022Dissecting373}. Furthermore, a large fraction, at least 30\%, of monitored deeply embedded protostars show observable rising/declining submm light curves stretching over many year timescales \citep{Lee2021TheProtostars,Mairs2024The2.0}. 

Within the JCMT Transient Survey sample, the Class 0 source HOPS 358 stands out as having the largest linear decline in 350 GHz (850\,$\mu$m) continuum brightness, of $\sim 25$\%, dropping continuously for four years before rising again \citep{Lee2021TheProtostars,Mairs2024The2.0}. The change in 
brightness observed by the JCMT scales approximately as the change in the envelope temperature and therefore the expected change in accretion luminosity (or equivalently, accretion rate) is expected to be much greater. Thus, HOPS 358 has likely under-gone a change in accretion luminosity of more than a factor of two
\citep[a reasonable expectation is $L \propto T^{\,>4}$; details of the theoretical and empirical scaling are given by][]{Johnstone2013ContinuumStructure,2020Contreras}.

HOPS 358 has previously been observed in the mid- and far-infrared by Herschel and classified as a, likely very young, PACs Bright Red Source by \citet{Stutz2013AProtostars} due to its large flux ratio between 70 and 24\,$\mu$m. Through spectral energy distribution fitting, \citet{Furlan2016THEMODELS} classified HOPS 358 as Class 0, with $T_{\rm bol} = 42$\,K and $L_{\rm bol} = 25\,L_\odot$. The environment around HOPS 358 has also been observed by ALMA as part of the ALMASOP survey of Orion \citep{Dutta2020ALMAOutflows}, where it is labelled G205.46-14.56S1\_A. Based on the diversity of the observed molecular lines, including complex organic molecules (COMs), the source is identified as a hot corino \citep{Hsu2022ALMACloud}. Furthermore, the outflow shows significant asymmetry, with SiO emission primarily evident in the red-shifted lobe. Regular clumping of material along the jet axis has been suggested \citep{Jhan2022ALMAKnots,Dutta2024ALMAProtostars}, which may indicate previous episodic accretion and ejection events.

Fortuitously, the VANDAM: Orion Survey of protostars in the Orion Molecular Cloud complex with ALMA \citep[e.g.][]{Tobin2020TheDisks} observed HOPS 358 over three epochs both before and near the start of its decline in brightness. Here we present an analysis of those observations along with six additional ALMA epochs 
of HOPS 358 to test whether the change in brightness observed with the JCMT with a $\sim$14\farcs6 beam (at $850\,\mu$m), tracing envelope scales, is reflected in the disk emission around the HOPS 358 protostar as probed by ALMA with a $\lesssim1$\arcsec\ beam. We also analyse the observed morphology of the disk to search for clues to any potential perturber.

In Section \ref{section:observations} we present details of the JCMT Transient Survey submm observations, NEOWISE mid-IR monitoring,  and our new ALMA observations of HOPS 358, along with the archival VANDAM: Orion ALMA data. In Section \ref{section:analysis} we present the variability of HOPS 358 observed by the JCMT Transient Survey before searching for variability across our nine epochs of ALMA observations. We also model our highest resolution ALMA observations to quantitatively measure properties of the disk around HOPS 358. Finally, in Section \ref{section:discussion}, we collect the information revealed by our multi-scale, multi-wavelength observations of HOPS 358 to investigate the origins of the variability seen towards HOPS 358.

\section{Observations \& Data Reduction}
\label{section:observations}

\subsection{JCMT}

The James Clerk Maxwell Telescope (JCMT), via the JCMT Transient Survey \citep{Herczeg2017HowRegions}, has been monitoring eight star-forming regions in the submm (450 and 850\,$\mu$m) using SCUBA-2 \citep{Holland2013SCUBA-2:Telescope} since December 2015. The observations are generally scheduled with an approximately monthly cadence. However, when HOPS-373 was seen to brighten in November 2019 \citep{Yoon2022Dissecting373} the NGC 2068 star-forming region, which includes HOPS 358, was boosted to a two week observing schedule, weather permitting.
Thus, as of August 2023 there have been 73 measurements of the brightness of HOPS 358 at 850\,$\mu$m (50 of which are sensitive at 450\,$\mu$m). SCUBA-2 observes at 850 and 450 $\mu$m simultaneously, however only when the weather is excellent are the 450\,$\mu$m measurements sensitive enough for brightness determinations \cite[see][]{Mairs2024The2.0}. In order to attain the desired cadence, the observations are therefore optimized for 850\,$\mu$m, with the shorter wavelength measurements, reliable only during the best weather, considered a bonus.

The JCMT Transient Survey has produced its own calibration algorithms  in order to robustly measure submm light curves \citep[for details see][]{Mairs2017TheMethods, Mairs2024The2.0}.  The calibration now achieves better than 1\% (5\%) relative brightness uncertainty at 850\,$\mu$m (450\,$\mu$m) for sufficiently bright sources, such as HOPS 358. For completeness, the effective beam sizes at the JCMT are 14\farcs6 and 9\farcs8 at 850\,$\mu$m and 450\,$\mu$m, respectively.

\begin{figure}
    \centering
    \includegraphics[width=3.4in]{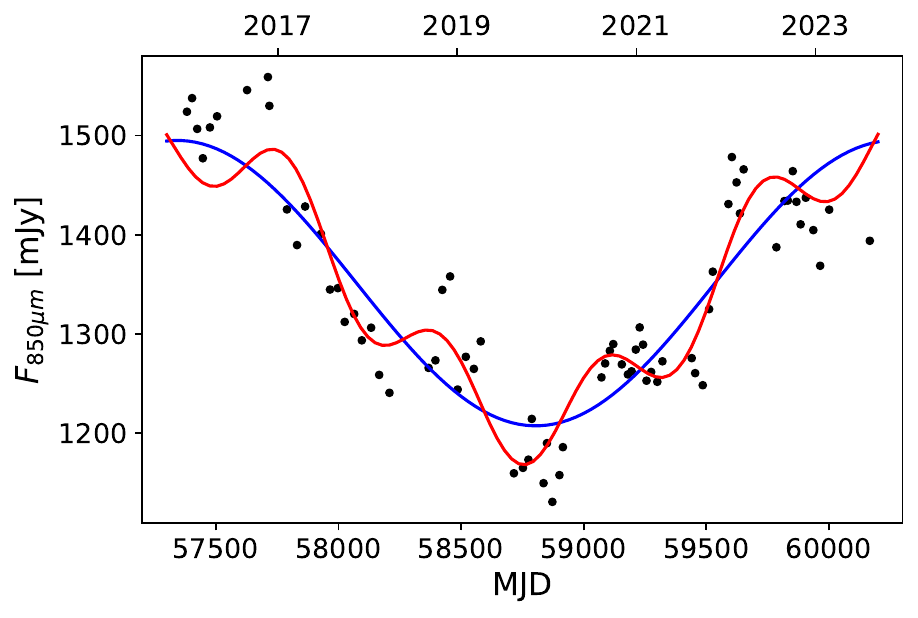}
    \vspace{-15pt}
    \caption{Light curve for HOPS 358 at 850\,$\mu$m (350 GHz) from the JCMT Transient Survey. Overlaid are single and double sinusoidal fits to the submm measurements (see text).}
    \label{fig:jcmt_850}
\end{figure}

The 850\,$\mu$m submm lightcurve for HOPS 358 is shown in Figure \ref{fig:jcmt_850} and Table \ref{table:jcmt_data}.
Between 2016 and 2024 the source has undergone a longer term dimming and recovery, modulated by a lower amplitude and shorter timescale periodicity.

\subsection{WISE/NEOWISE}

In this work we use mid-infrared (mid-IR) photometry from all-sky observations of the {\it WISE} space telescope. {\it WISE} surveyed the entire sky in four bands, W1 (3.4 $\mu$m), W2 (4.6 $\mu$m), W3 (12 $\mu$m), and W4 (22 $\mu$m), with the spatial resolutions of 6\farcs1, 6\farcs4, 6\farcs5, and 12\arcsec, respectively, from January to September of 2010 \citep{Wright2010THEPERFORMANCE}. The survey continued as the NEOWISE Post-Cryogenic Mission, using only the W1 and W2 bands, for an additional 4 months \citep{Mainzer2011NEOWISERESULTS}. In September 2013, {\it WISE} was reactivated as the NEOWISE-reactivation mission \citep[NEOWISE-R,][]{Mainzer2014INITIALMISSION}. {NEOWISE-R operated until mid-2024, and the latest released data set contains observations until mid-December 2023.} HOPS 358 is observed every six months, with each visit consisting of
several photometric observations with a few day window. 

For HOPS 358, we used all the available data observed between 2010 and 2022. The single-epoch data was collected from the NASA/IPAC Infrared Science Archive (IRSA) catalogues \citep{irsa144} using a 10\arcsec~radius from the coordinates of the YSO. We averaged the single epoch data taken a few days apart to produce 1 epoch of photometry every 6 months \citep[following the procedures described in][]{Park2021QuantifyingExplorer}.

\subsection{ALMA}

We observed HOPS 358 with ALMA across six epochs from 21 October 2019 until 03 January 2022 in two separate Programs (2019.1.00691.S, 2021.1.00844.S, PI: Sheehan). As our observations were spread out over two years, the configuration of the array changed significantly over the course of our six epochs. In Table \ref{table:observations} we provide a summary of our observations, including observation date, calibrators used, range of baselines observed, and synthesized beam sizes.

\begin{deluxetable*}{lcllcc}
\tablecaption{Summary of ALMA Observations}
\tablenum{2}
\tabletypesize{\scriptsize}
\label{table:observations}
\tablehead{\colhead{Date} & \colhead{Baselines} & \colhead{Calibrators} & \colhead{Self-calibration Intervals} & \colhead{Beam Size} & \colhead{RMS}\\ \colhead{ } & \colhead{k$\lambda$} & \colhead{flux,bandpass,phase} & \colhead{Attempted/{\bf Applied}} & \colhead{$\mathrm{{}^{\prime\prime}}$} & \colhead{$\mathrm{\mu Jy}$}}
\startdata
2016-09-03 & $  13 - 1998$ & \nodata & \nodata & $0.19 \times 0.13$ & 587 \\
2016-09-04 & $  16 - 2728$ & \nodata & \nodata & $0.13 \times 0.11$ & 520 \\
2017-07-19 & $  18 - 3592$ & \nodata & \nodata & $0.14 \times 0.09$ & 479 \\
2019-10-21 & $  16 -  839$ & J0510+1800,J0510+1800,J0552+0313 & {\bf inf\_EB},96.77s,48.38s,30.24s,12.10s,{\bf int},{\bf inf\_ap} & $0.39 \times 0.31$ & 249 \\
2021-07-04 & $  32 - 3062$ & J0510+1800,J0510+1800,J0552+0313 & {\bf inf\_EB},inf,26.21s,{\bf 12.10s},6.05s,int,300s\_ap,inf\_ap & $0.12 \times 0.08$ & 121 \\
2021-08-22 & $  47 - 12872$ & J0510+1800,J0510+1800,J0552+0313 & {\bf inf\_EB},{\bf inf},26.21s,12.10s,6.05s,int,{\bf inf\_ap} & $0.03 \times 0.02$ & 32 \\
2021-11-03 & $  44 - 5666$ & J0510+1800,J0510+1800,J0552+0313 & {\bf inf\_EB},{\bf inf},26.21s,12.10s,6.05s,int,{\bf 300s\_ap},inf\_ap & $0.06 \times 0.05$ & 112 \\
2022-01-03 & $  16 -  824$ & J0510+1800,J0510+1800,J0552+0313 & {\bf inf\_EB},96.77s,48.38s,30.24s,12.10s,{\bf int},{\bf inf\_ap} & $0.34 \times 0.30$ & 324 \\
2021-12-30 & $  10 -  882$ & J0725-0054,J0538-4405,J0541-0541 & {\bf inf\_EB},96.77s,48.38s,30.24s,12.10s,{\bf int},inf\_ap & $0.54 \times 0.27$ & 429
\enddata
\end{deluxetable*}

The observations were designed to match the spectral set up from the VANDAM: Orion observations of HOPS 358 \citep{Tobin2020TheDisks} in order to facilitate a consistent comparison between these new epochs 
and the existing archival observations from 2016 and 2017. Our spectral setup therefore utilized four spectral windows centered at 330.887965, 333.0, 334.0, and 345.79599 GHz with bandwidths of 0.9375, 1.875, 1.875, 0.9375 GHz and channel widths of 0.564, 1.129, 1.129, and 0.564 MHz (0.511, 1.016, 0.984, and 0.489 km/s velocity resolution). We note that the archival observations by \citet{Tobin2020TheDisks} placed the continuum spectral windows at 333.0 and 334.0 GHz in TDM continuum mode, however we set up those corresponding windows in our observations in FDM mode with higher spectral resolution to search for spectral lines that happen to fall within the bandpass. We also note that the archival observations utilized a narrower bandwidth, of 0.234 GHz, for the spw centered at 330.887965 GHz.

The observations were initially calibrated, including bandpass-, gain-, and flux-calibration using the calibrators listed in Table \ref{table:observations}, using the ALMA Interferometric Pipeline \citep{Hunter2023TheHeuristics}. The absolute fluxes, in particular as this paper is concerned with temporal brightness variations, were calibrated by including in each observing track observations of a quasar whose flux is monitored regularly by ALMA to use as a reference, and comparing the flux of that source measured in that track with the monitored flux. The expected accuracy of the absolute flux calibration from this procedure is 10\% for our Band 7 observations, meaning that all fluxes {directly} extracted from our observations will have an additional 10\% uncertainty on their values in addition to the statistical uncertainties from the flux extraction.\footnote{See, e.g., the ALMA Proposers Guide here:\\ https://almascience.nrao.edu/proposing/learn-more}

\begin{figure*}
    \centering
    \includegraphics[width=7in]{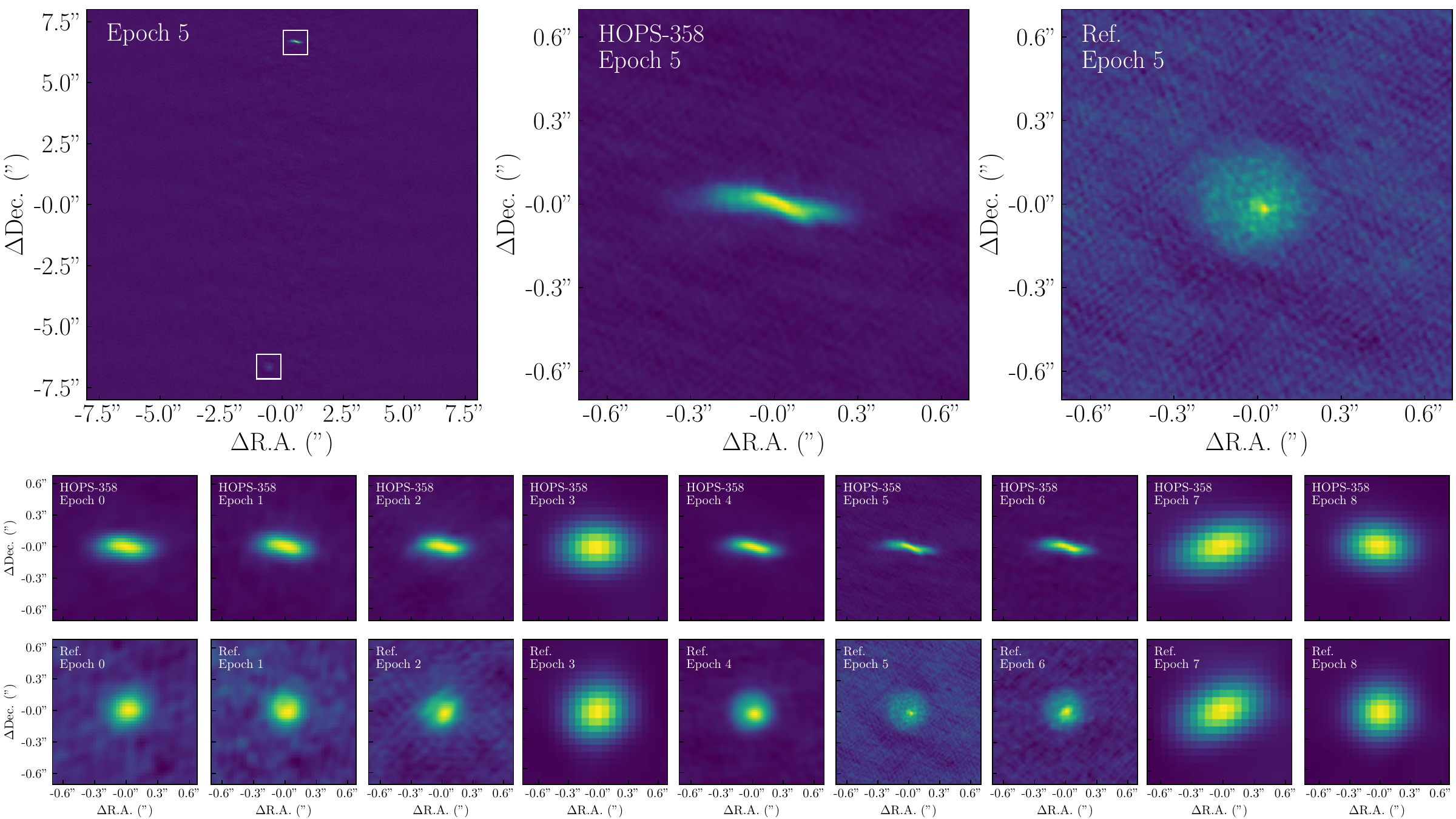}
    \caption{({\it top row}) Our highest resolution images of the full field of view showing HOPS 358 and HOPS 358B separated by $\sim13"$ ({\it left}), HOPS 358 ({\it center}), and HOPS 358B ({\it right}). ({\it bottom two rows}) Images of both HOPS 358 and HOPS 358B from each epoch of our observations, at the native resolution of those data.
    }
    \label{fig:alma_images}
\end{figure*}

As HOPS 358 is a bright source, we then followed up the standard calibrations by self-calibrating each epoch. To perform self-calibration, we used the development version of the standalone version of the ALMA Pipeline self-calibration pipeline (Tobin et al., in prep.)\footnote{Available here: github.com/psheehan/auto\_selfcal}. In short, multiple rounds of self-calibration were performed, CLEANing more deeply and shortening the solution interval for the gain calibration solutions at each iteration. The first iteration solves for calibrations per-spw, per-polarization over the entire track, i.e. with {\it combine=``scan"} and {\it solint=``inf"}, and is pre-applied before solving for subsequent intervals with {\it combine`=`spw"}. After phase-only self-calibration, the routine also attempts amplitude self-calibration, pre-applying the phase-only self-calibration tables that were previously derived. We list the self-calibration solution intervals attempted and ultimately accepted in Table \ref{table:observations}.

We produce initial images of each epoch of our observations using the {\it tclean} task within CASA \citep{TheCASATeam2022CASAAstronomy} using multi-frequency synthesis for the continuum images and Briggs weighting with a robust parameter of 0.5. We list the beam shape (major and minor axes and position angle) and RMS of each epoch in Table \ref{table:observations}, and show the continuum images in Figure \ref{fig:alma_images}. Our observations span a range of spatial resolutions, but the highest resolution observations reveal an edge-on disk with a clear warped structure.

In addition to HOPS 358, our images also show a companion source $\sim13"$ to the south. Prior observations of this target are limited \citep[e.g.][]{Chen2013SMASYSTEMS,Dutta2020ALMAOutflows,Galametz2020AnProtostars}, though the lack of an infrared detection across a range of observations from $1 - 160$ $\mu$m leads \citep[e.g.][]{Dutta2020ALMAOutflows} to suggest that it is a particularly young, Class 0 protostar. Polarized emission is detected in a region offset to the west of the millimeter peak of this {more southern} source \citep[e.g.][]{Galametz2020AnProtostars}. Though the origin of this emission is unclear, \citet{Luo2022ALMAProtostars} shows a collimated outflow traced by $^{12}$CO with approximately the same orientation. Our highest angular resolution observations of this source show a remarkably spherical structure with a point-like source offset slightly south and west of its center. We discuss the nature of this source further in Section \ref{section:hops358s}, but it is additionally important as a potential reference source to use to examine the variability of HOPS 358. We henceforth refer to this source as HOPS 358B.

\subsection{Archival VANDAM: Orion Observations}

In addition to our new ALMA observations, we collect archival observations of HOPS 358 at 345 GHz taken as a part of the VANDAM: Orion survey \citep{Tobin2020TheDisks}. Details of the observations and their calibration are described in detail by \citet{Tobin2020TheDisks}, but we do include the relevant details in Table \ref{table:observations} such that that information is consolidated into one location here. We further include continuum images from those observations in Figure \ref{fig:alma_images}. 

\section{Analysis \& Results}
\label{section:analysis}

\subsection{Variability on Envelope Scales}

\begin{figure*}[t]
    \centering
    \includegraphics[width=7in]{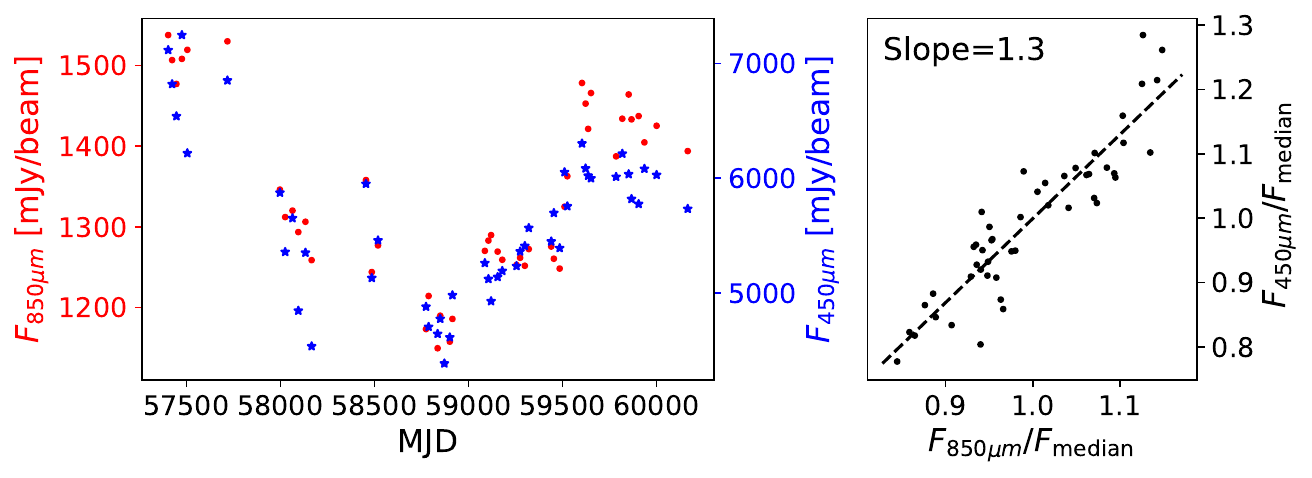}
    \vspace{-13pt}
    \caption{(Left) Combined JCMT light curve for HOPS 358 at 450 and 850\,$\mu$m, for data points observed simultaneously. (Right) Comparison between the 450 and 850\,$\mu$m variable response, with each normalized by the median value. Note that the 450\,$\mu$m measurements show a significantly stronger degree of variability due to their probing closer to the peak of the envelope spectral energy distribution (see text).}
    \label{fig:jcmt_850_450}
\end{figure*}

The 850\,$\mu$m submm lightcurve for HOPS 358 (Figure \ref{fig:jcmt_850}) reveals a strong decline and recovery over the eight year monitoring duration. The best-fit sinusoidal curve to these measurements suggests a period of about eight years, though it should be noted that fitting to a single cycle is fraught with uncertainty, including the appropriateness of the applied fitting function.  
{Indeed, given that the sinusoidal fit assumes an unobserved turn-over before the first measurement and after the last, we stress that the eight year dipping event timescale is only a lower limit and that there is no evidence of repetition.}
Nevertheless, the long period sinusoid does a very good job at smoothly following the observed lightcurve, with an amplitude of $\sim$10\% the mean submm brightness and a peak to peak brightness variation of $\sim$20\%.
On removal of the long timescale variability, a clear shorter period residual
{with multiple cycles} is revealed. Thus, a combined two sinusoid fit uncovers a second period of 1.75\,yr with an amplitude of $\sim$3\% the mean submm brightness of the source. The fit parameters, including the false alarm probability of each fit, are provided in Table \ref{table:jcmt_sin_fit}.

\begin{deluxetable}{lcc}
\tablecaption{JCMT 850$\mu$m Sinusoidal Fits}
\tablenum{3}
\tabletypesize{\normalsize}
\label{table:jcmt_sin_fit}
\tablehead{\colhead{Parameter}        & \colhead{First} & \colhead{Second} }
\startdata
Mean flux (mJy beam$^{-1}$) & 1351.3 & 1.75 \\
Period\tablenotemark{\scriptsize a} (yr)& 8.01& 1.75 \\
Amplitude\tablenotemark{\scriptsize a} (mJy beam$^{-1}$)\  & 143.9 & 42.7 \\
Residual RMS (mJy beam$^{-1}$) & 49.6 & 40.0 \\
False Alarm Probability\tablenotemark{\scriptsize a} & $5\times10^{-24}$ & $6\times10^{-6}$ \\
\enddata
\tablenotetext{a}{Derived from Lomb-Scargle periodogram fitting \citep{lomb1976, scargle1989} .}
\end{deluxetable}

Assuming that the submm 
{variability} observed in HOPS 358 is due to the thermal reprocessing of an underlying variable accretion luminosity by the optically thick and submm thin dust envelope, there should be a measurable change in the amplitude of the observed variability with changing submm wavelength due to stronger deviations from a Rayleigh-Jeans temperature-only response at wavelengths closer to the peak of the emitted spectral energy distribution, as described by \citet{Johnstone2013ContinuumStructure}. Figure \ref{fig:jcmt_850_450} compares the HOPS 358 light curves at both 450 and 850\,$\mu$m, where measurements were taken simultaneously. There is an apparent additional 30\% stretch in the variability at 450\,$\mu$m compared with the 850\,$\mu$m observations. This quantification is consistent with expectations, as discussed by \citet{2020Contreras} and \citet{Mairs2024The2.0}, assuming an outer envelope dust temperature, $T_d \sim 20$\,K.

\begin{figure}
    \centering
    \includegraphics[width=3.4in]{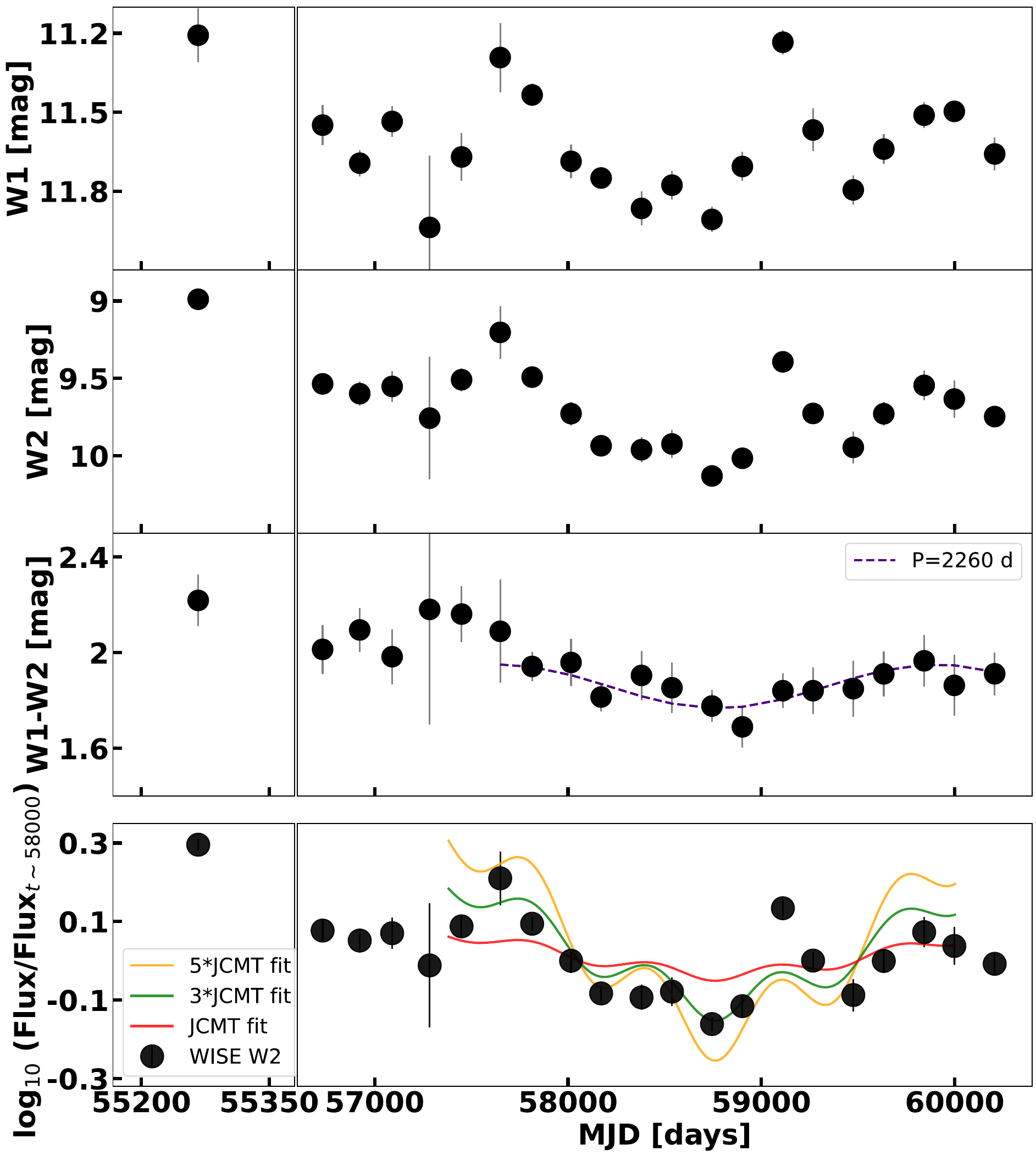}
    \includegraphics[width=3.4in]{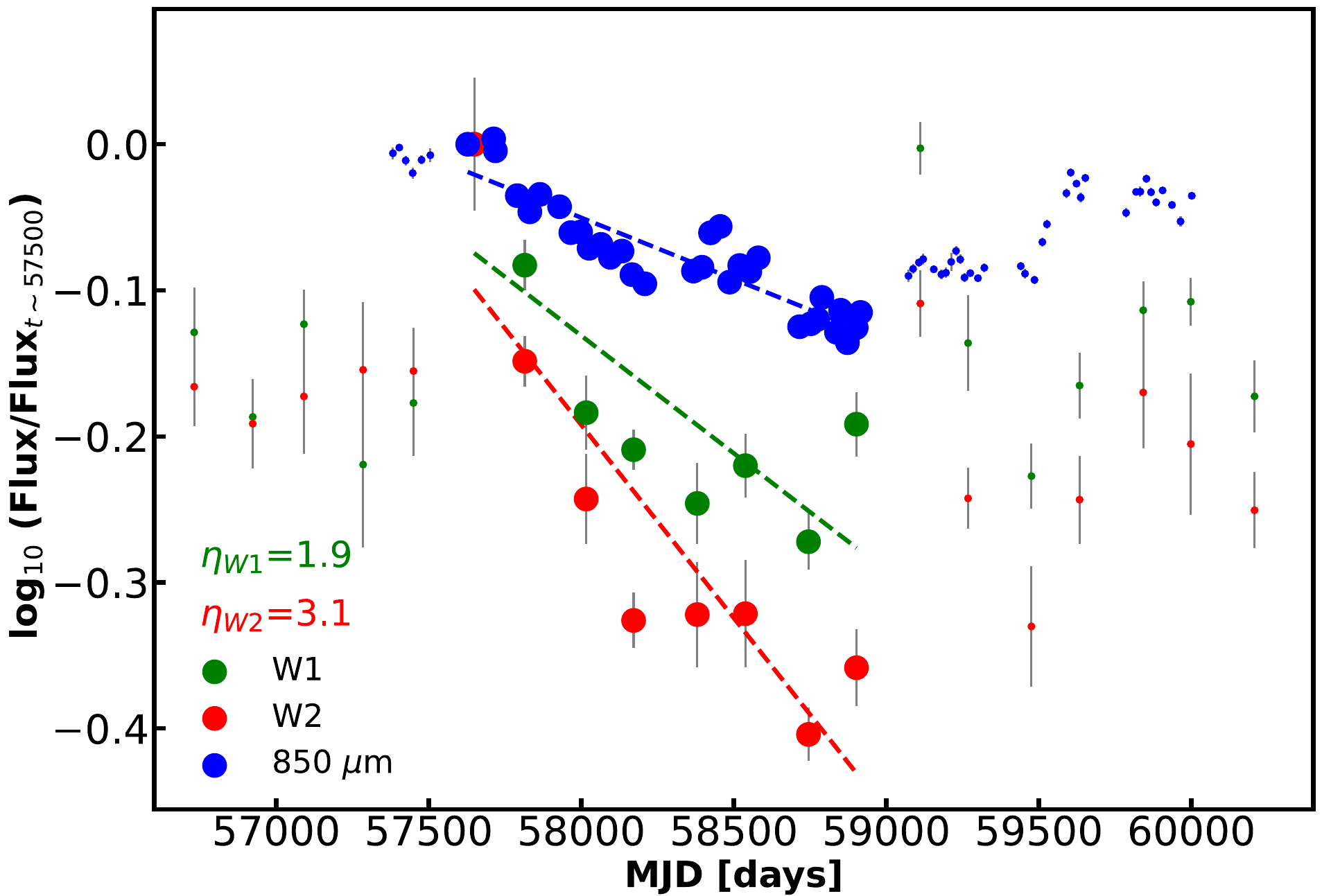}
    \caption{{(top)  WISE W1, W2, and W1-W2 light curves of HOPS358. The W1-W2 colours that are contemporaneous with the JCMT observations are fitted by a sinusoidal model with P=2260 d (indigo line). (middle)  Log of the flux normalized at MJD$\sim$58000 d for the WISE W2 (black circles) data of HOPS 358. In the same figure we show the fit to the JCMT 850 $\mu$m flux (red line in Fig. \ref{fig:jcmt_850}), and the same fit but using scaling factors of 3 (green solid line) and 5 (orange solid line). (bottom) Log of the flux for the 850 $\mu$m, WISE W1 and W2 observations. For all of these, we fit the region between 57500 d$<$MJD$<$59000 d (large circles) with a linear model. This provides values of the correlation, $\eta$ \citep[see][]{2020Contreras}, between mid-IR and sub-mm fluxes. We find $\eta_{W1}=1.9$ and $\eta_{W2}=3.1$  }}
    \label{fig:midir}
\end{figure}

{The mid-IR W1, W2, and W1$-$W2 light curves
of HOPS 358 are presented in Figure \ref{fig:midir}. The W1 and W2 light curves show high-amplitude ($\Delta \sim 1$~mag) variability across the 12 years of WISE/NEOWISE coverage. 
These amplitudes are fairly consistent with the expectations from changes in the accretion rate within such a system \citep[e.g.][]{Scholz2013APhotometry}.
Further, these light curves suggest two timescales. Localized bursts have $\sim$year durations and overlie what appears to be a longer, many years, secular timescale. Interestingly, the W1$-$W2 light curve shows significantly less of the short term variability, making the longer term secular evolution more apparent.}

\begin{figure*}[t]
    \centering
    \includegraphics[width=7in]{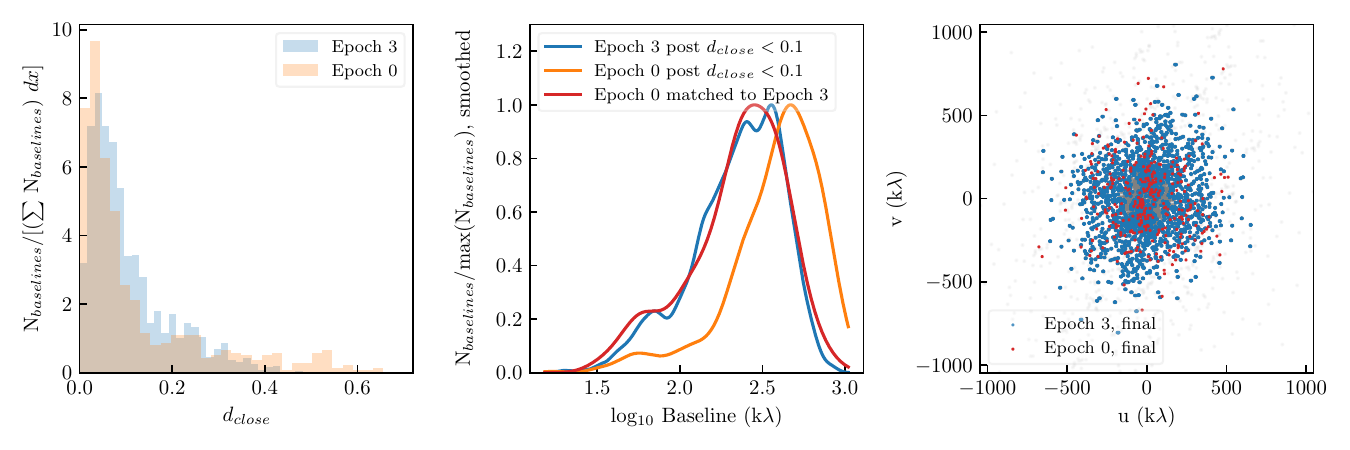}
    \vspace{-10pt}
    \caption{Demonstration of the process of matching baselines between two epochs of observations. On the left we show the histograms of $d_{close}$  calculated for each epoch, {shown normalized by total number of baselines times the bin size to keep datasets with large differences in the number of baselines on the same scale}. In the center we show KDE estimates of the distribution of baselines for each epoch of data after the cut of $d_{close} < 0.1$, along with an addition distribution showing the baselines for the epoch matched to Epoch 3. Finally, on the right we show the final, two-dimensional distribution of baselines for the two epochs after this matching routine. We also show all of the trimmed points as faint transparent gray points in the background. The complete figure set (8 images) is available in the online journal.}
    \label{fig:baseline-matching}
\end{figure*}

{Considering only the date range encompassing both WISE/NEOWISE and JCMT coverage, bottom panel of Figure \ref{fig:midir}, the mid-IR W2 observations can be reasonably reproduced using an amplitude-scaled version of the sub-mm two-component light curve model. The observed W2 variability is demonstrably stronger than the unscaled sub-mm model (red line), as scaling factors between 3 and 5 (green and orange) provide reasonable, by eye, fits. The short, $\sim$year, mid-IR bursts, however, are not well matched by the sub-mm model and there is no obvious connection between their occurrence and the phase of the shorter period component of the sub-mm model. Strong short timescale mid-IR variability without matching sub-mm variability is not unexpected, as the mid-IR probes a small region near the protostar which reacts quickly to accretion variability, while the single-dish sub-mm emission comes from the much larger envelope and only responds slowly to accretion luminosity variations \citep[see][]{Francis2022AccretionAssembly}.}

{{Given that the sub-mm and mid-IR observations are not taken at the same time, and there are large amplitude peaks with durations of $\sim$1 year in the mid-IR not apparent in the sub-mm, it is difficult to find an exact scaling factor \citep[$\eta$ in][]{2020Contreras}. We derive a 
quantitative 
estimate by fitting linear models to the mid-IR and sub-mm fluxes observed in the range 57500$<$MJD$<$59000 d (see bottom panel of Fig. \ref{fig:midir}). The comparison of the slope of the fits provides values of $\eta$=1.9 and 3.1 when comparing the sub-mm with the W1 and W2 fluxes, respectively.  The top end of the 2 -- 5 range discussed above from linear fits and visual inspection, however, is consistent with the typical values found across an ensemble of variable protostars by \citet{2020Contreras}. There, a 5 times scaling was shown to be roughly that expected from continuum radiative transfer through the dense protostellar envelope. We caution that contamination of the continuum emission by non-varying molecular emission in the mid-IR bands \citep[c.f. HOPS 373][]{2022Yoon} can lower the observed scaling factor.}}

{{Finally, in further support of the supposition that the mid-IR emission is reacting in a similar manner to the sub-mm, we note that the mid-IR color, W1$-$W2, undergoes an apparent secular variation quite similar to the longer period sub-mm fit, with less stochasticity than the individual W1 and W2 light curves. We fit a sinousoidal model to the region of the W1-W2 light curve that is contemporaneous with the JCMT observations. A curve with $P=2260$ d provides a reasonable fit to the data. Therefore, the mid-IR color does not suggest a longer timescale than the $\sim$8 year fit to the sub-mm observations.}
During the brighter sub-mm epochs the mid-IR color tends to shift towards larger values of W1$-$W2. This tendency to become redder when brighter has been observed in other young eruptive protostars such as HOPS 267, LDN 1455 IRS3 \citep{2023Contreras}, SPICY 978455 \citep{2023Contreras_b}, EC53 \citep{Lee2020YoungDisk}, HOPS12 and HOPS124 \citep{Zakri2022TheOrion}, and WISEA J142238.82-611553.7 \citep{Lucas2020DiscoveryAmplitude}.}

\begin{figure*}
    \centering
    \includegraphics[width=7in]{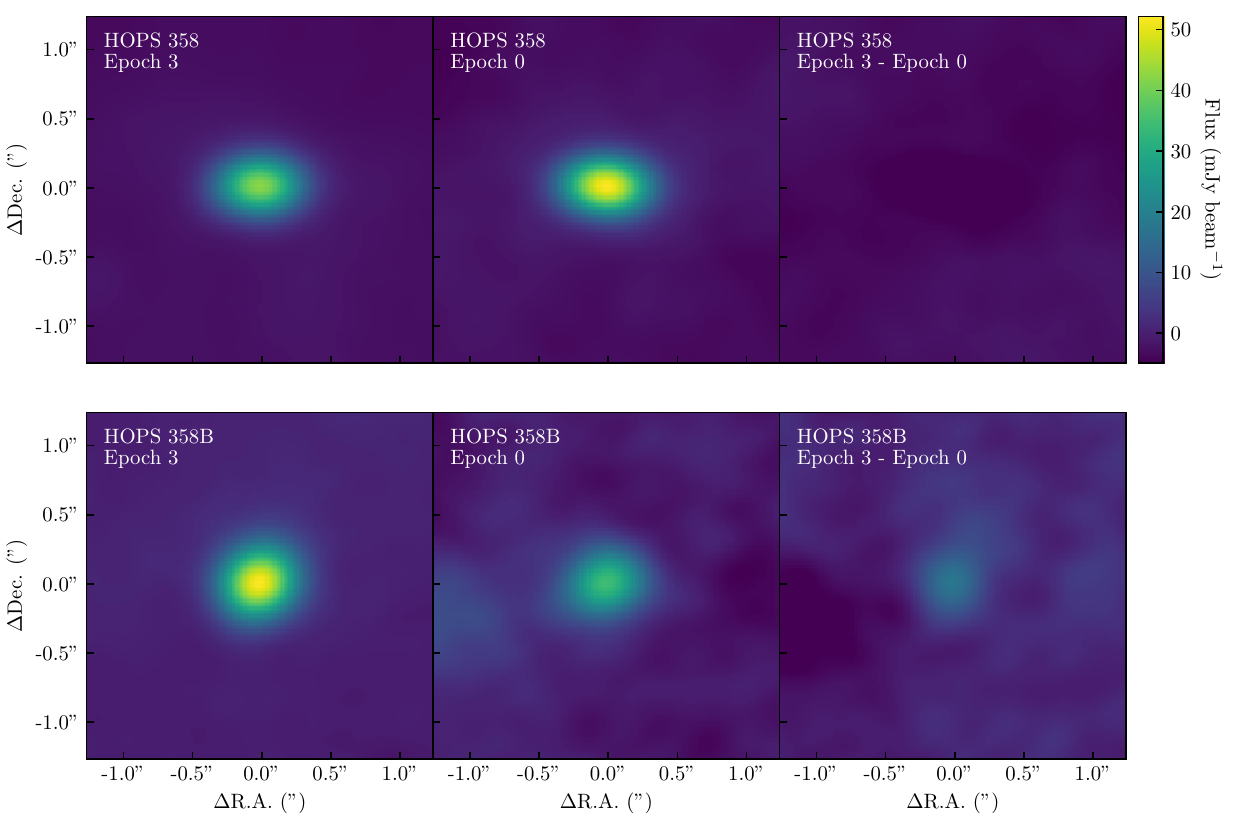}
    \caption{Comparison of HOPS 358 ({\it top}) and HOPS 358B ({\it bottom}) for each epoch of data with our reference epoch, Epoch 3, when imaged using only baselines matched to Epoch 3 as described in Section \ref{section:alma-flux-comparison}. Though we show the images as well as the difference image between them here for a visual demonstration, our quantitative comparison of the flux between epochs is done by fitting a two-dimensional Gaussian function in the visibility plane to each of the matched datasets. We also note that we have not attempted to correct for uncertainties in the relative flux scaling between epochs, so the peak fluxes and structures in the residuals should be treated with caution. Relative changes between HOPS-358 and HOPS-358B are, however, more reliable. The complete figure set (8 images) is available in the online journal.
    }
    \label{fig:matched-images}
\end{figure*}

\subsection{Flux Comparison Across ALMA Epochs}
\label{section:alma-flux-comparison}

As our observations span a range of different ALMA observing configurations, comparing the flux of a given source in said observations from epoch to epoch is a non-trivial endeavor. Interferometric images cannot be directly compared as even in the event that the observing configuration remained the same, observations at different times will have unique sampling, and therefore differing beams, in the $uv$-plane due to the Earth's rotation. To mitigate these effects and compare fluxes for HOPS 358 across epochs, we therefore follow an adaptation of the technique proposed by \citet{Francis2019IdentifyingCARMA}. In short, the goal of this procedure is to ``similarize" the $uv$-plane coverage across observations to ensure that the data being compared are sensitive to similar structures.

To do this for a pair of observations, following \citep{Francis2019IdentifyingCARMA} we first calculate the distance from a given observed baseline in the {observation of interest} to the nearest observed baseline in the {reference} observation for every observed baseline in the {observation of interest}, and vice-versa. We normalize these distances by the distance from zero in the uv-plane so that the normalization scales with increasing baseline; i.e. at longer baselines the absolute distance to the nearest point is relaxed compared to at shorter baselines. For both datasets, we remove any baselines for which the fractional distance, $d_{close} > 0.1$.  We found that this fraction provides a reasonable balance between not including measurements that are too far from any baselines in the other dataset, but also not throwing away too much data (e.g.\ left panel of Figure \ref{fig:baseline-matching}).

\begin{figure*}
    \centering
    \includegraphics[width=7in]{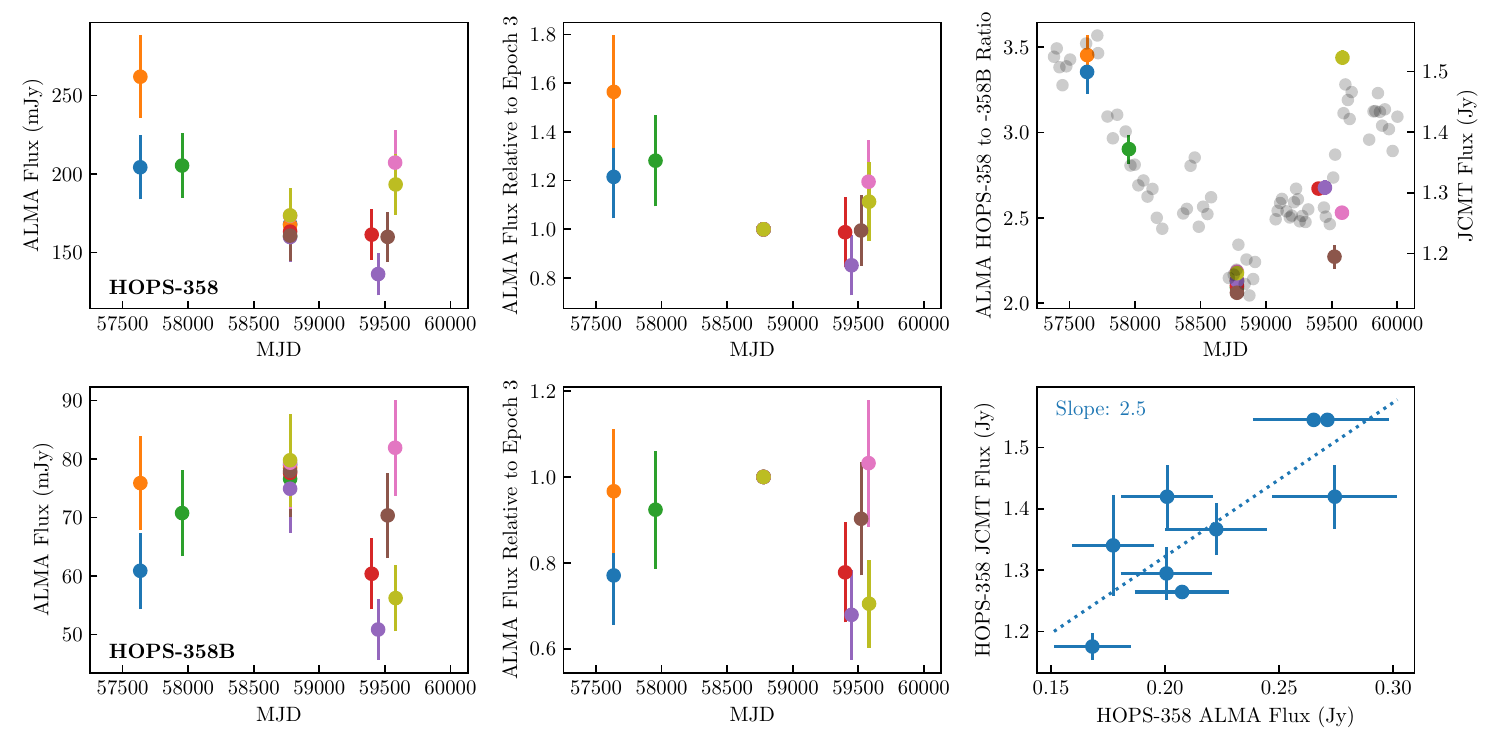}
    \vspace{-10pt}
    \caption{Comparison of the flux of HOPS 358 ({\it top}) and HOPS 358B ({\it bottom}) across our nine epochs of data. In the left column we show the absolute flux measured from our data after matching the baselines with Epoch 3 as described in Section \ref{section:alma-flux-comparison}, noting that for Epoch 3 there is a measurement for comparison with each individual Epoch. In the center column we show the flux measured for each epoch of data relative to the flux measured in Epoch 3. {In the top row of the} the third column we show the ratio of the flux of HOPS 358 to the flux of HOPS 358B for each epoch of our data, with the JCMT 850 $\mu$m fluxes in grayscale for comparison. {Finally, in the bottom row of the third column, we show a comparison of the HOPS 358 ALMA flux with the JCMT flux at the same epoch, calculated as an average of the JCMT fluxes measured within 90 days of the ALMA epoch. We further show a linear fit to the data, and include the slope of that fit, as well.}}
    \label{fig:flux-comps}
\end{figure*}

At this point, each dataset should only contain measurements that are at a similar location in the $uv$-plane to data in the {paired} dataset. Differing array configurations of ALMA, however, means that the weighting of the data in the radial direction of the $uv$-plane can still differ substantially (e.g. the center panel of Figure \ref{fig:baseline-matching}). For the example shown in Figure \ref{fig:baseline-matching}, even though we have discarded baselines that are too far from baselines in the {paired} dataset, the distribution of baselines as a function of radius peaks to longer baselines for Epoch 0 than for Epoch 3 because Epoch 0 was observed in a more extended configuration. To remedy this imbalance, we use a Kernel Density Estimator to produce a smoothed distribution of baselines for each epoch, normalized to a peak of one, and then take the ratio of the {reference epoch} to the  {epoch of interest}. This ratio of smoothed baseline distributions is calculated at every individual baseline in the {epoch of interest} and used as a probability threshold for whether to keep that data point or to drop it. In dropping the data points that are not accepted, we end up with a radial distribution of baselines for the {epoch of interest} that approximately matches the distribution of baselines for the {reference} epoch (see e.g. the central panel of Figure \ref{fig:baseline-matching}).

With this sampling of baselines from the {epoch of interest}, we end up with baselines for both epochs that approximately match in distribution, and therefore should produce images with approximately similar beams. That said, the process is not entirely perfect at matching beams, so rather than comparing images directly with, {for example} a subtraction or peak flux measurement, we instead fit a two dimensional Gaussian function to the remaining datasets from both epochs to measure source properties, such as total flux. In doing so, we can be confident that the scales and structures being probed by the Gaussian fit are similar. For ease of viewing, however, we do show a comparison of the images generated from the remaining baselines from both epochs as well as a difference image in Figure \ref{fig:matched-images}.

Finally, as our ALMA observations span a large range in spatial scales, and as Figure \ref{fig:alma_images} demonstrates, HOPS 358's disk ranges from marginally resolved to very well resolved. Indeed, from our longer baseline datasets it is clear that a two dimensional Gaussian is not a good fit to the data. We therefore use Epoch 3, which has the lowest resolution, $0.39" \times 0.31"$, of all of our datasets and in which HOPS 358 is only marginally resolved, as the reference epoch, and compare all other epochs to this one. For each epoch we fit two Gaussian components, one to represent HOPS 358 and the other to represent HOPS 358B. The fluxes measured from these fits are reported in Table \ref{table:best_fits}

We show a comparison of the beam-normalized fluxes measured for HOPS 358 and HOPS 358B across our 9 epochs in Figure \ref{fig:flux-comps}. We show these fluxes in three different ways: the left and center panels show the absolute fluxes and the absolute fluxes normalized to Epoch 3, respectively, for HOPS 358 and the HOPS 358B. In these panels, we note that we have included a 10\% flux calibration uncertainty in addition to the statistical uncertainty from the model fit. From looking at the absolute fluxes of HOPS 358 and HOPS 358B with time, we find that while there is some variation in the flux, and indeed the lightcurve for HOPS 358 qualitatively follows a similar trend as is seen from both the JCMT as well as the mid-infrared, the 10\% flux calibration uncertainty renders it impossible to determine whether these differences are statistically significant.

In order to overcome the flux calibration uncertainty, in the third panel of Figure \ref{fig:flux-comps} we show the ratio of the flux of HOPS 358 to HOPS 358B. As this is a relative quantity, it should be unaffected by the flux calibration uncertainty and provide an absolute quantity that can be compared epoch to epoch. The difficulty, however, is that it can only tell us whether {\it something} in the field of view is changing but cannot decide which (or both) of the two sources is varying. We find that there is a statistically significant decline in the ratio through our fourth epoch, followed by a slight rise in the ratio of brightness during the final few epochs.

For comparison, we also show the JCMT observations compared with the ratio of HOPS 358 to HOPS 358B in the third column of Figure \ref{fig:flux-comps}. The overall trend, a decreasing flux until MJD $\sim58750$ followed by a rise, matches quite well between the two datasets. The good match in the overall pattern lends significant credence to the idea that it is variations of HOPS 358 that is driving the overall trend seen in the ALMA data.

While the overall trend is similar between the JCMT fluxes and the ALMA flux ratios, the magnitude of the variations for the ALMA data {is larger than the magnitude of the variations seen by the JCMT. Indeed, in the lower right corner of Figure \ref{fig:flux-comps} we show a plot of the JCMT fluxes against the ALMA fluxes and find a linear trend with a slope of 2.5. That is to say that the variations in the JCMT flux are only $2.5\times$ larger than the variations in the ALMA flux despite the JCMT measuring $\sim6\times$ as much flux. The difference is even larger if the disk contribution to the JCMT flux, as measured by ALMA, is removed, leaving just the flux from the envelope. In that scenario, the change in envelope flux is only a factor of $\sim1.5\times$ larger than the variations in the disk flux in spite of being $\sim5\times$ brighter. It is worth noting, though, that the change in brightness of the disk alone cannot explain the change in brightness seen by the JCMT.}

There are some statistically significant variations in the ratio of brightness of the two sources during the final few epochs that happen on relatively short timescales, most notably the large rise seen between Epochs 8 \& 9, which are separated by four days.  
The light-crossing time for HOPS 358's disk, with a radius of $\sim0\farcs25 = 100$ au at a distance of $\sim$400 pc \citep[e.g.][]{Kounkel2017THECLOUDS,Zucker2019AEdition} is 0.57 days, and the timescale for the dust to respond to the increased luminosity is short \citep[e.g.][]{Johnstone2013ContinuumStructure}. So it is conceivable that these short timescale variations seen in the ratio of brightness could be attributed to the disk around HOPS 358. Indeed, considering the known variability of HOPS 358 on longer timescales, this {is} the simplest explanation for these short timescale variations. Such short-term changes in the brightness would be washed out in the JCMT observations by the considerably larger beam
{ \citep{Francis2022AccretionAssembly}}. 
Still, as we only have HOPS 358B for reference, it cannot be ruled out that this southern companion is also varying, and might dominate the short-term variations.

\begin{deluxetable*}{cccccccc}
\tablecaption{Fluxes}
\tablenum{4}
\tabletypesize{\small}
\label{table:best_fits}
\tablehead{\colhead{Epoch} & \colhead{MJD} & \colhead{HOPS-358 Flux} & \colhead{HOPS-358 Flux,} & \colhead{HOPS 358B Flux} & \colhead{HOPS 358B Flux,} & \colhead{HOPS 358-to-358 B} & \colhead{HOPS 358-to-358 B} \\ \colhead{} & \colhead{} & \colhead{} & \colhead{Ref. Epoch} & \colhead{} & \colhead{Ref. Epoch} & \colhead{Ratio} & \colhead{Ratio, Ref. Epoch}  \\ \colhead{} & \colhead{} & \colhead{(mJy)} & \colhead{(mJy)} & \colhead{(mJy)} & \colhead{(mJy)} & \colhead{} & \colhead{}}
\startdata
0 & 57634.6 & $202.4 \pm {2.0}$ & $168.08 \pm {0.43}$ & $60.6 \pm {2.2}$ & $79.02 \pm {0.46}$ & $3.34 \pm {0.13}$ & $2.127 \pm {0.013}$ \\
1 & 57635.5 & $262.1 \pm {2.1}$ & $167.49 \pm {0.41}$ & $75.9 \pm {2.5}$ & $78.52 \pm {0.44}$ & $3.45 \pm {0.13}$ & $2.133 \pm {0.013}$ \\
2 & 57953.7 & $205.4 \pm {1.9}$ & $160.24 \pm {0.43}$ & $70.8 \pm {2.0}$ & $76.62 \pm {0.47}$ & $2.902 \pm {0.086}$ & $2.091 \pm {0.014}$ \\
4 & 59399.6 & $161.34 \pm {0.81}$ & $163.31 \pm {0.41}$ & $60.40 \pm {0.81}$ & $77.66 \pm {0.43}$ & $2.671 \pm {0.038}$ & $2.103 \pm {0.014}$ \\
5 & 59448.5 & $136.17 \pm {0.77}$ & $159.81 \pm {0.55}$ & $50.87 \pm {0.86}$ & $74.94 \pm {0.61}$ & $2.677 \pm {0.046}$ & $2.132 \pm {0.019}$ \\
6 & 59521.3 & $159.9 \pm {1.8}$ & $160.72 \pm {0.50}$ & $70.4 \pm {2.0}$ & $77.98 \pm {0.55}$ & $2.272 \pm {0.071}$ & $2.061 \pm {0.016}$ \\
7 & 59578.3 & $207.30 \pm {0.59}$ & $173.39 \pm {0.38}$ & $81.94 \pm {0.60}$ & $79.39 \pm {0.39}$ & $2.530 \pm {0.019}$ & $2.184 \pm {0.012}$ \\
8 & 59582.1 & $193.35 \pm {0.72}$ & $173.62 \pm {0.37}$ & $56.24 \pm {0.69}$ & $79.81 \pm {0.39}$ & $3.438 \pm {0.044}$ & $2.176 \pm {0.012}$
\enddata
\tablecomments{Epoch 3 is used as the  reference epoch (``Ref. Epoch"). Columns marked as such in this table show the flux measured for the relevant source in the reference Epoch using the baselines recovered after matching the spatial frequency coverage as described in Section \ref{section:alma-flux-comparison}}
\end{deluxetable*}

\subsection{Disk Structure from Analytic Modeling}

Our high resolution ALMA Band 7 imaging of HOPS 358 shows that the disk is both large ($\sim0.5" = \sim200$ au in diameter) and warped. To better characterize the properties of this structure, we fit analytic models to the brightness profile of the disk seen in our long baseline dataset to quantify the disk properties such as radius, radius of the warp, and total brightness.

We use as our initial model brightness profile a simple rectangle model with smoothed edges.  Previously, \citet{Sheehan2022AIRS} used such a model to fit the edge-on disk around L1527. The intensity profile is given by
\begin{equation}
    \label{equation:rectangle}
    I_{r} = I_0 \, \exp\left\{ -\frac{(x - x_c)^{\gamma_x}}{2 \, x_w^{\gamma_x}} - \frac{(y - y_c)^{\gamma_y}}{2 \, y_w^{\gamma_y}}\right\},
\end{equation}
where $(x_c,y_c)$ is the center of the rectangle along the major and minor axes, $\gamma_x$ and $\gamma_y$ control how sharply the rectangle is truncated along the major and minor axes, and $x_w$ and $y_w$ control the width of the disk along the major and minor axes.

We fix the disk origin at the protostar, therefore, $x_c = 0$; to account for the warping of the disk, however, rather than centering the Gaussian for the rectangle along the minor axis at $y_c = 0$, we instead modify the equation for a smoothly broken power-law such that $y_c = 0$ for $x \ll x_{b}$ and $y_c$ linearly increases with $x$ for $x \gg x_{b}$, with $x_b$ denoting the location of the break along the major axis, and the two solutions are smoothly joined:
\begin{equation}
    y_{c} = (x_w - x_{b}) \, \tan{\theta_{b}} \, \left(\left(0.5 \, \left(1 + \frac{|x|}{x_{b}}^{1/\Delta}\right)\right)^{\Delta} - 0.5^{\Delta}\right).
\end{equation}
Though perhaps not immediately obvious, for $x \ll x_b$ the term in parentheses reduces to $y_{c} \approx (x_w - x_b) \, \tan{\theta_b} \, \left(0.5^{\Delta} - 0.5^{\Delta}\right) = 0$. For $x \gg x_b$ and $\Delta \ll 1$ (a prior on our model), this becomes $y_{c} = (x_w - x_b) \, \tan{\theta_b} \, \, (\frac{|x|}{x_b} - 1)$, which is a line with slope $(x_w - x_b) \, \tan{\theta_b}$. As such, $\theta_b$ can be interpreted as the angular break in the disk and $\Delta$ controls the break sharpness.

\begin{figure*}
    \centering
    \includegraphics[width=7in]{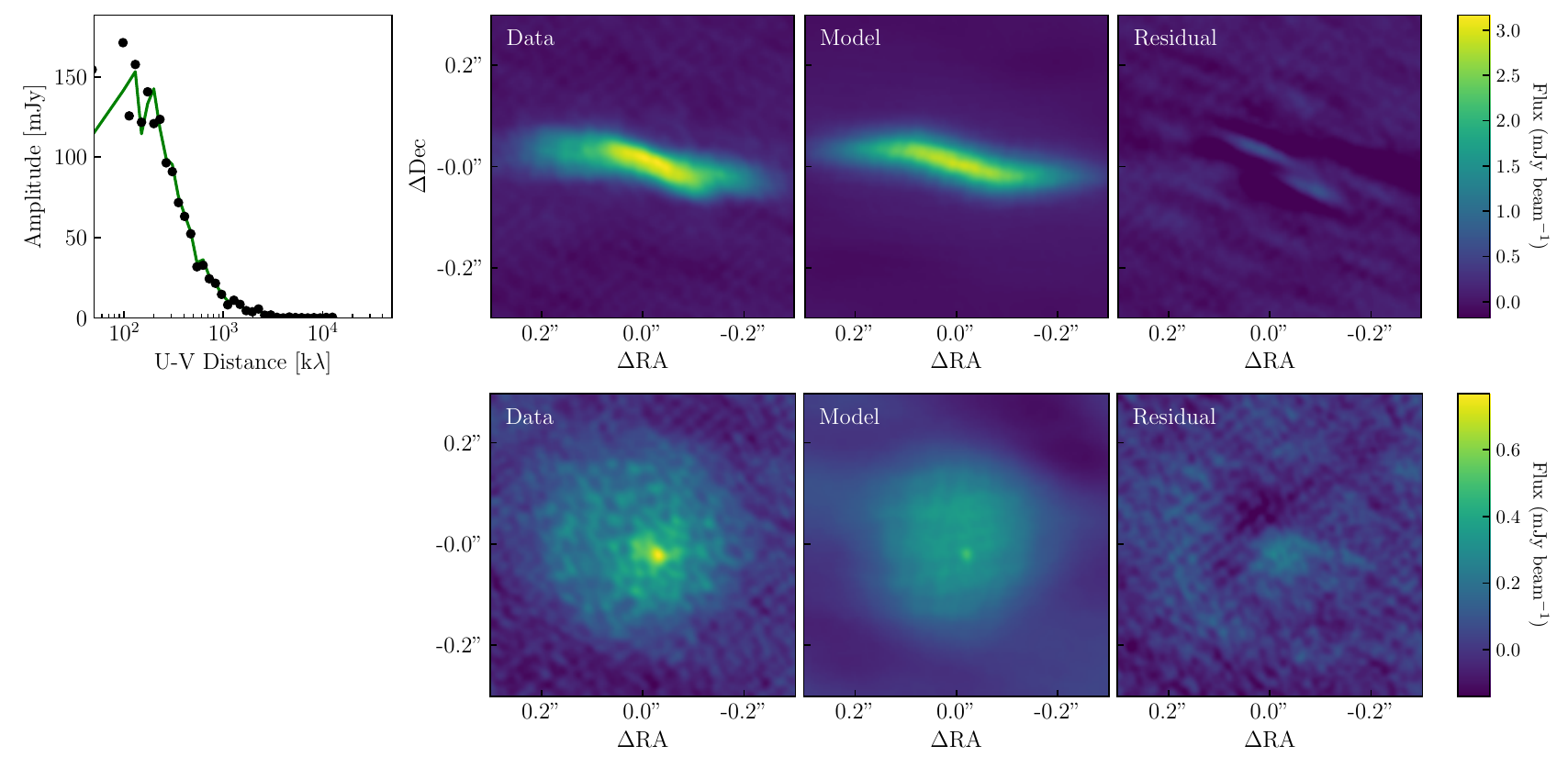}
    \caption{Best-fit analytic model for our high resolution ALMA imaging of the disk around the HOPS 358 protostar. In the leftmost column we show the one-dimensional, azimuthally averaged visibility profile for HOPS 358 ({\it black}) with the best-fit model, generated at the same visibility points as our data and averaged in the same way, shown ({\it green}) for comparison. In the remaining three columns we show data, model and residual plots for both HOPS 358 ({\it top}) and {HOPS 358B} ({\it bottom}). The model and residual images are generated by Fourier-transforming and deconvolving (in the case of the model image) the model/residual visibilities in the same way as was done for the data.
    }
    \label{fig:high-res-model}
\end{figure*}

With the model defined, we then shift the phase center from $(0,0)$ to the observed phase center of HOPS 358 at $(x_0,y_0)$, where $x_0$ and $y_0$ are left as free parameters, and rotate the model by some position angle to match the data. 
In addition, rather than specifying $I_0$, the peak model intensity, we use the total intensity, $F_{\nu,disk}$ as a free parameter. To scale the model to have the correct flux, we generate the model image with $I_0 = 1$, integrate this model image to calculate the total flux in that image, and scale the image by the ratio of $F_{\nu,disk}$ to the total flux in the model image with $I_0 = 1$.

To ensure that the structure of HOPS 358B is not affecting our modeling, we also include a simple model for it. We start with a circle with smoothed edges,
\begin{equation}
    I_{south} = I_0 \exp\left\{ -\frac{r^{\gamma_c}}{2 \, r_w^{\gamma_c}}\right\}.
\end{equation}
To account for the faint point source that appears slightly to the south-west of the center of the circle, we additionally add a point source to this model at some offset $(x_p,y_p)$ from the center of the circle. As with the disk model, rather than leaving the peak intensity ($I_0$) as a free parameter, we integrate over the entire circle model and use the total integrated flux $F_{\nu,circle}$ as the free parameter. We keep the flux of the point source component $F_{\nu,point}$ as a free parameter of the model as well. Like with the disk, we shift the phase center of the entire circle+point model to the phase center of HOPS 358 in the data, $(x_s,y_s)$, where $x_s$ and $y_s$ are left as free parameters.

We separately generate model images for the warped disk and circle, and use the \texttt{galario} \citep{Tazzari2018GALARIO:Observations} code to Fourier transform each image into the visibility plane at the observed baselines of our observations and shift them to the appropriate phase centers. As point source models are trivial to generate directly in the visibility plane, we skip the image for the point source component and directly generate model visibilities. The visibilities for the three components are then added together to produce the final model visibilities.

To fit the model to the data and find the best-fit parameters and uncertainties, we use the \texttt{dynesty} code \citep{Speagle2020Dynesty:Evidences}, which employs the Nested Sampling algorithm, to sample parameter space. We report the best-fit parameters in Table \ref{table:analytic_best_fits} and show the best-fit model compared with the data in Figure \ref{fig:high-res-model}. We define the best-fit model as the maximum-likelihood model from our Nested Sampling fitting, and the uncertainties listed are derived from the range around the best-fit model that includes 68\% of the posterior samples.

We find that the best-fit disk model does a reasonable job of reproducing our high resolution observations of HOPS 358. Though there remain some significant residuals, likely due to the complex structure of the disk, the fit can reproduce the overall features seen in the image. Per the best-fit model, the disk has a radius, as defined by $x_w$ in our model, of $0.1779\farcs$, corresponding to $\sim71$ au at a distance of $\sim$400 pc. The break in the disk occurs at $0.088\farcs$, or $\sim35$ au, and the disk breaks at an angle of $16^{\circ}$. The best-fit model also finds that the circular component of HOPS 358B has a radius of $0.1277\farcs$, corresponding to 51 au at a distance of $\sim$400 pc. We do find that the point-like source is significantly offset from the center of this circular component, with an offset vector of ($-0.022\farcs$,$-0.023\farcs$), or ($-8.8$ au, $-9.2$ au) at $\sim$400 pc, with positive offsets being in the East and North directions.

\begin{deluxetable*}{cc|c|cc|c}
\tablecaption{Best-fit Analytic Model Parameters}
\tablenum{5}
\tabletypesize{\normalsize}
\label{table:analytic_best_fits}
\tablehead{\colhead{Parameter} & \colhead{Unit} & \colhead{HOPS-358} & \colhead{Parameter} & \colhead{Unit1} & \colhead{HOPS-358}}
\startdata
$x_0$ & $''$ & $0.0140^{+0.0003}_{-0.0003}$ & $p.a.$ & $^{\circ}$ & $253.4^{+0.2}_{-0.1}$ \\[2pt]
$y_0$ & $''$ & $0.00698^{+0.00009}_{-0.00012}$ & $F_{\nu}$ & mJy & $146.7^{+0.4}_{-0.3}$ \\[2pt]
$x_w$ & $''$ & $0.1779^{+0.0003}_{-0.0004}$ & $x_c$ & $''$ & $1.1243^{+0.0011}_{-0.0008}$ \\[2pt]
$y_w$ & $''$ & $0.0221^{+0.0005}_{-0.0003}$ & $y_c$ & $''$ & $-13.2705^{+0.0009}_{-0.0008}$ \\[2pt]
$x_b$ & $''$ & $0.088^{+0.002}_{-0.001}$ & $r_w$ & $''$ & $0.1277^{+0.0004}_{-0.0008}$ \\[2pt]
$\theta_{b}$ & $^{\circ}$ & $16.3^{+1.0}_{-1.4}$ & $\gamma_c$ & \nodata & $2.80^{+0.03}_{-0.06}$ \\[2pt]
$\Delta$ & \nodata & $0.10^{+0.02}_{-0.07}$ & $F_{\nu,c}$ & mJy & $38.7^{+0.3}_{-0.3}$ \\[2pt]
$\gamma_x$ & \nodata & $2.51^{+0.01}_{-0.02}$ & $x_g$ & $''$ & $-0.022^{+0.005}_{-0.006}$ \\[2pt]
$\gamma_y$ & \nodata & $0.74^{+0.02}_{-0.01}$ & $y_g$ & $''$ & $-0.023^{+0.005}_{-0.004}$ \\[2pt]
$\sigma$ & \nodata & $0.0001^{+0.0007}_{-0.0001}$ & $F_{\nu,g}$ & mJy & $0.30^{+0.03}_{-0.05}$
\enddata
\end{deluxetable*}

\section{Discussion}
\label{section:discussion}

To summarize the results of Section \ref{section:analysis}, we find that HOPS 358 is variable at submm wavelengths, as evidenced by the JCMT Transient Survey observations, as well as at mid-infrared wavelengths, observed by NEOWISE. Moreover, the JCMT Transient survey measurements suggest that there exists 
two timescales associated with the envelope scale variability, the shorter of which appears quasi-periodic.
Our ALMA observations demonstrate that something in the data is varying, but cannot conclusively attribute that variability to either HOPS 358 or its southern companion due to the sole use of HOPS 358B as a point of reference. Assuming that HOPS 358B does not vary, then the compact disk emission around HOPS 358 seen by ALMA reacts similarly to the dominant long-timescale JCMT envelope-scale measurement, albeit with a somewhat larger amplitude. Furthermore, ALMA reveals that the protostellar disk around HOPS 358 is warped, with a $16^{\circ}$ warp at $\sim35$ au. 

{The larger amplitude of fluctuations seen by ALMA than seen by the JCMT could plausibly be explained in a number of ways. One possibility is that there is a stronger temperature variation in the disk, perhaps because the disk is not solely heated passively by the protostar but is rather heated by the accretion processes directly \citep{Liu2022DiagnosingNear-infrared}. {While the energy liberated in the disk would not add appreciably to the passive heating of the envelope, it would locally affect the temperature of the disk}. Alternatively, it is possible if not probable that the envelope is heated not only by the protostar, but also by external radiation, which would dampen the impact of variations in heating by the protostar
\citep{Johnstone2013ContinuumStructure, Francis2022AccretionAssembly,Baek2020RadiativeObject}. The radiative transfer of the system may further play a role here; some fraction of the envelope is likely shadowed by the disk, which is optically thick at short wavelengths. The disk may then absorb and reemit at longer wavelengths that can escape the system many of the photons that might have otherwise heated the system. How impactful such an effect might be, however, would be highly dependent on the geometry (i.e.~scale height, flaring) of the disk, and require detailed radiative transfer modeling beyond the scope of this paper.}

Three {further} questions arise from these observations: (1) what physical mechanisms are responsible for the warp, (2) what physical mechanisms are responsible for the observed variability time-scales and amplitudes, and (3) are there any connections between the observed warp and the measured variability of HOPS 358. Indeed it is tempting to attribute the warp and the variability to the same underlying physical mechanism, {even though the driver of variability is traditionally attributed to gravitational instabilities in disks \citep[e.g.][]{Vorobyov2007Self-regulatedDiscs,Zhu2009TWO-DIMENSIONALOUTBURSTS} or internal disk physics such as magneto-rotational instabilities (MRI) \citep{Bae2014ACCRETIONDISKS, Flock20173DDisks}, whereas warps are frequently associated with unseen companions \citep[e.g.][]{Nealon2018WarpingOrbit,Zhu2019InclinedSignatures,Facchini2018SignaturesObservations}.} Furthermore, the period of a Keplerian orbit at 35 au around a 1 M$_{\odot}$ star is $\sim200$ years, which is dramatically different from the eight years found for the longer of the variability timescales obtained from  our single-dish submm observations. This directly observed timescale instead suggests a Keplerian orbit origin would be around a few au. While the Keplerian timescale does scale weakly with protostellar mass ($P \propto M^{-0.5}$), HOPS 358 would need to have an unrealistically high mass in order to bring these timescales into alignment.  
It is worth reiterating that there is considerable uncertainty on the estimated timescale of the large amplitude variability seen by the JCMT
{and whether it repeats in a regular manner,} 
given the lack of completion of a full cycle. 
{Nevertheless,} extending the cycle to a century or longer would either require an extreme amplitude or a non-sinusoidal shape.

To further complicate the timescale issue, the ALMASOP team has studied the outflow associated with HOPS 358 (also known as G205.46-14.56S1\_A) using ALMA, revealing the presence of knots of material spaced at regular intervals of $\sim1\farcs$ \citep[][]{Jhan2022ALMAKnots,Dutta2024ALMAProtostars}. Such knots in a jet have previously been suggested as signposts of accretion bursts onto the central protostar \citep[e.g.][]{Plunkett2015EpisodicSouth} that would drive variability. 
According to \citet{Jhan2022ALMAKnots} and \citet{Dutta2024ALMAProtostars}, the period of whatever is inciting the knots is only a few decades, 20-35\,yr, though there is considerable uncertainty to longer timescales given the difficulty of determining the true jet velocity due to the almost plane of sky outflow inclination angle. The decades period is not that far off from the $\sim8$ years 
{dip event that} we measure with the JCMT, recognizing the previous caveats.

We further note that, though the $\sim25\%$ decrease in submm brightness may ostensibly seem too small to drive such features in an outflow, the emission traces primarily the temperature variations \citep[e.g.][]{Johnstone2013ContinuumStructure}, and the underlying change in luminosity, driven in turn by the change in accretion, is likely much larger. \citet{2020Contreras} found a relationship between the {protostellar luminosity} and millimeter brightness {such that ${F_{850\,\mu\textrm{m}}} \propto L^{0.27}$, so the increase in the protostellar luminosity, and therefore the accretion rate, for HOPS 358 is likely on the order of a factor of $\sim2.3$.} This is also reasonably consistent with the $\sim1$ mag of variability seen in the WISE/NEOWISE data for HOPS 358 (e.g. see Figure \ref{fig:midir}).

\subsection{Disk Gravitational Instability}

The common explanation for variability in the luminosity of protostars is that as the disk grows in mass the disk becomes graviationally unstable. Traditionally the disk mass grows due to a mismatch between accretion from the infalling envelope \citep[e.g.][]{Ulrich1976AnPhenomenon,Terebey1984TheClouds} and the steady-state rate of flow through the disk and onto the star, though in principle disk mass can build-up at localized radii as material flows through the disk \citep[e.g.][]{Pinilla2012TrappingDisks}. Induced gravitational instabilities facilitate the transport of angular momentum \citep[e.g.][]{Kratter2016GravitationalDisks} and thereby drive a period of rapid accretion onto the central protostar \citep[e.g.][]{Vorobyov2007Self-regulatedDiscs,Zhu2009TWO-DIMENSIONALOUTBURSTS,Zhu2010Long-termOutbursts,Bae2014ACCRETIONDISKS}, thereby increasing its luminosity.

To test whether gravitational instabilities could be present within the HOPS 358 disk, we use the best-fit parameters from our physical model to estimate the gravitational stability of the disk using the Toomre Q parameter,
\begin{equation}
    Q = \frac{c_s \, \Omega}{\pi \, G \, \Sigma},
\end{equation}
{where $c_s$ is the sound speed, $\Omega$ is the angular rotational speed, and $\Sigma$ is the surface density. $Q < 1$ is the traditional limit for gravitational instabilities, though somewhat higher values might still be considered marginally unstable and capable of transporting angular momentum and driving spiral arms \citep[e.g.][]{Kratter2010TheStars}.}
To estimate the surface density of the disk $\Sigma$, we first calculate the total mass of the disk by assuming that the disk is optically thin such that,
\begin{equation}
    M_d = \frac{d^2 \, F_{\nu}}{\kappa_{\nu} \, B_{\nu}(T)}.
\end{equation}
As HOPS 358's disk is edge-on, it is quite likely that it is actually at least partially optically thick. So the above equation likely produces a lower limit on the true mass of the disk, though how far off is difficult to characterize without more detailed modeling.
\citet{Tobin2020TheDisks} studied a grid of radiative transfer models and found that this equation best reproduced the known disk masses of the model grid if the disk-averaged temperature above was scaled by the protostellar luminosity $L_*$,
\begin{equation}
    T = 43 \, \mathrm{K} \, \left(\frac{L_*}{1 \, L_{\odot}}\right)^{0.25}.
\end{equation}
We use the above equation with the measured bolometric luminosity of 25 L$_{\odot}$ \citep{Tobin2020TheDisks} to calculate a temperature of 96 K, and use the flux for the disk measured from our best-fit model in Table \ref{table:analytic_best_fits}{, a dust opacity of 3.45 cm$^2$ g$^{-1}$ \citep[e.g.][]{Beckwith1990AObjects}}, and a gas-to-dust ratio of 100 to find a total disk mass of 0.01 M$_{\odot}$. 

We then assume that this material is spread evenly in a uniform disk with a radius of 70 au, as given by our best-fit model, such that the surface density of the disk is a constant value given by the ratio of the disk mass to the area of the disk. Though we note that this method almost certainly does not match the true surface density distribution of the disk, it does give a reasonable, order-of-magnitude estimate for the purposes of assessing the disks stability.

We assume the protostar has a central mass of 1 M$_{\odot}$ to calculate the Keplerian velocities, and use the bolometric luminosity of 25 L$_{\odot}$ to calculate the temperature as a function of radius from
\begin{equation}
    T = \left(\frac{L_*}{4 \pi \sigma r^2}\right)^{0.5}.
\end{equation}
As the temperature and velocity both decrease with radius, and we have assumed a constant surface density, Q will be lowest at the outer edge of the disk. We therefore calculate Q at the disk radius of 70 au, with the understanding that it will likely be even higher further into the disk interior. Under these assumptions, we find that the disk has $Q\sim17$ at the outer edge.

Such large values of $Q$ {would suggest} that the disk is stable against self-gravity \citep[e.g.][]{Kratter2010TheStars}. Even if we assume a lower luminosity of $11$ L$_{\odot}$ prior to the outburst, based on the factor of 2.3 estimated previously, we still find $Q\sim15$ at the outer edge.
{On the other hand, this calculation is also sensitive to choice of $M_*$ and dust opacity. If, assuming a dust opacity of 1.84 cm$^2$ g$^{-1}$ \citep[e.g.][]{Ossenkopf1994DustCores,Tobin2020TheDisks} or a protostar mass of 0.25 M$_{\odot}$ each yields Q values $\sim2\times$ lower, or a factor of $4\times$ lower together. In aggregate, this would produce $Q\sim4.5$, closer to the value of $1.6$ that \citet{Tobin2020TheDisks} finds, and also closer to the range where the disk may be at least marginally gravitationally unstable. Optical depth could also play a role; indeed, the disk is edge-on and likely at least somewhat optically thick at 345 GHz. \citet{Tobin2020TheDisks} find a disk mass that is $\sim3\times$ higher using the VLA flux, which would further reduce the Q value (in \citet{Tobin2020TheDisks} this yields $Q = 0.7$, in the range where the disk is likely gravitationally unstable), however the VLA emission is also more compact and could be impacted by free-free emission and so should be treated with caution. Still, it is conceivable that gravitational instabilities could play a role in the HOPS-358A disk. Estimates of the protostellar mass as well as better estimates of the disk mass may help to remove some of these uncertainties and better characterize the stability of this disk.}

\subsection{Other Disk Instabilities}

As noted in the accretion variability review by \citet{Fischer2023AccretionAssembly}, there are many potential disk instabilities operating at various distances from the central protostar which might account for years-long observed brightness variability. These include convective and or thermal instabilities \citep{Kadam2019DynamicalFormation, Pavlyuchenkov2020EvolutionRegions}, which can provide both intermittent episodic accretion events and quasi-periodic variable accretion, and instabilities associated with MRI-induced interchange of energy and angular momentum within the disk leading to efficient inward mass transport \citep{Zhu2010Long-termOutbursts,Bae2014ACCRETIONDISKS, Flock20173DDisks}. In particular, the MRI mechanism for accretion is expected to work well in the outer disk, where cosmic rays can maintain the required disk ionization level for effective coupling with the magnetic field, and in the very inner disk, where ionizing radiation from the central source can again induce sufficient ionization. In the intermediate regions of the disk, however, the MRI may not couple well as the disk becomes more neutral producing a dead-zone. This situation creates the appropriate conditions for accretion-related instabilities, either at the inner edge of the dead-zone, yielding short-timescale events and potentially producing quasi-periodic feedback, or from the outer dead-zone edge, invoking longer timescale runaway accretion episodes.

We note that the shorter timescale observed for HOPS 358, 1.75\,yr, is similar in duration to the 1.5\,yr quasi-periodic variability timescale found for EC\,53 in Serpens \citep{Hodapp2011ERUPTIVENW,Lee2020YoungDisk}. For EC\,53 this timescale was analysed, using an $\alpha$-disk model, and associated with a potential disk blockage at a very small radii $\sim$0.05\,au \citep{Lee2020YoungDisk}. Interestingly, similar to EC\,53, this short timescale associated with HOPS 358 shows clear 
{repeatability} and thus maybe the same underlying physical accretion instability is at play, perhaps associated with the inner-edge of an MRI dead-zone or a thermal instability. The significant difference between the two sub-mm variable sources is that where the short timescale periodicity dominates in EC\,53, for HOPS 358 it appears to be much weaker in amplitude.

\subsection{Origins of the Disk Warp}

Whichever disk instability is driving protostellar accretion, it is not immediately clear how to include the warp of the disk without invoking additional physics{,} though we do not rule out that the warp and the variability could still be linked in some way, perhaps with the warp providing a ringing of the inner disk and therefore driving local shorter timescale disk instabilities{.}
One possible mechanism for producing a warp is continued interactions of the system with its natal environment. Indeed, simulations of the collapse of giant molecular clouds to form protostars and their disks \citep[e.g.][]{Bate2018OnDiscs,Kuffmeier2021MisalignedInfall} show how disks might continue to interact with their environments during the course of their lifetime. For example, \citet{Kuffmeier2021MisalignedInfall} shows an example of a disk that accretes its initial disk, and then later accretes additional material with a different angular momentum vector, leading to misaligned disks that might otherwise appear warped. 
Alternatively, simulations of stellar fly-by's have also suggested that such an interaction could induce a warp in a disk that persists for thousands of years \citep[e.g.][]{Cuello2019FlybysDynamics}.
 
As mentioned earlier, another common explanation for disk warps is the presence of an unseen companion, interacting gravitationally with the disk.
Simulations have demonstrated that a planet embedded in a disk with an orbit misaligned with the disk can drive a warp between the disk interior to the planet and the disk exterior to the planet \citep[e.g.][]{Nealon2018WarpingOrbit,Zhu2019InclinedSignatures}. Alternatively, it has also been demonstrated that a binary system whose orbit is misaligned with the orbit of its circumbinary disk can also drive a warp in the inner disk \citep[e.g.][]{Facchini2018SignaturesObservations,Bi2020GWAction}. 
{Though we stress that we have no direct evidence for the presence of such a companion, the latter scenario of a misaligned binary could simultaneously explain the warp and the 
{variability}
seen in the JCMT observations.}

\subsection{Speculation on HOPS 358B}
\label{section:hops358s}

Though we have thus far focused primarily on HOPS 358, the southern companion HOPS 358B shows interesting structure, in its own right. The lack of infrared detection across a range of observations from $1 - 160$ $\mu$m \citep[e.g.][]{Dutta2020ALMAOutflows} and presence of a collimated outflow \citep[e.g.][]{Luo2022ALMAProtostars} would seem to suggest that it is a particularly young protostar. {The 870 $\mu$m and 9 mm emission seen towards HOPS 358B is reminiscient of the emission towards the sample of particularly young protostars presented by \citet{Karnath2020DetectionFormation}, with spatially resolved, low-contrast emission at 870 $\mu$m giving way to much higher contrast, point-like, emission at 9 mm where optical depth is lower \citep[see Figure 25 of][]{Tobin2020TheDisks}. The emission towards HOPS 358B, however, is much more symmetric at 870 $\mu$m than most of the sample from \citet{Karnath2020DetectionFormation}, with the exception of the off-center central peak.}

While one might typically associate the relatively compact{,} $<0\farcs4$ in diameter{,} millimeter emission associated with this source with a disk \citep[e.g.][]{Tobin2020TheDisks}, we find that the peak is significantly offset from the center of the circular emission. Though this might be indicative of asymmetries in the disk, we do not see evidence of such features in our high resolution image, although that image is relatively noisy. Alternatively, the recent eDisk survey demonstrated  that there are asymmetries along the minor axis of many young disks due to inclination and optical depth effects \citep[e.g.][]{Ohashi2023EarlyResults}. The offset peak we see towards HOPS 358B could have a similar origin, though the near circular appearance that would tend to imply a closer to face-on orientation might make this more challenging. Optical depth effects from a particularly dense envelope with a possible outflow cavity could conceivably contribute to the appearance as well.

\section{Conclusions}

To conclude, we have analyzed nine epochs of ALMA observations at sub-arcsecond resolution of the protostar HOPS 358 and its companion HOPS 358B located $13"$ to the south during a period of variability identified by the JCMT Transient Survey on $\sim15"$ scales. Our main conclusions are as follows:
\begin{itemize}
    \item HOPS 358 is observed to vary in the JCMT Transient Survey observations, with a decline in brightness of $\sim25\%$ over forty months before beginning to rise again.
    \item The amplitude and color variation of the mid-IR photometry of HOPS 358 is similar to that of known eruptive YSOs. The light curve follows a similar pattern to the observed variability at sub-mm wavelengths, supporting the interpretation as an accretion-driven outburst \citep[see][]{2020Contreras}. The scaling of the mid-IR to sub-mm fluxes is fairly consistent with the results from \citet{2020Contreras}.
    \item If we assume that the variability observed is periodic and sinusoidal then we estimate a period of 8 years, and residuals from the single-sinusoid fit suggest a second period of about 1.75 yr. It is important to note the considerable uncertainty in the longer 
    {timescale measurement, however, as we have not yet observed any repetition.}
    \item Commensurate ALMA observations show a time-varying ratio of the brightness of HOPS 358 {relative} to its southern companion, HOPS 358B, that matches the trend seen in the JCMT observations remarkably well. With that said, attributing the trend seen in the ALMA observations entirely to variability of HOPS 358 requires assuming that HOPS 358B is, itself, not varying.
    \item In addition to brightness variability of the disk, our highest resolution ALMA observations demonstrate the HOPS 358's disk is warped, with a $16^{\circ}$ warp of the disk at about 35 au, halfway through the disk.
    \item The traditional explanation for protostellar variability, of gravitational instabilities driving changes to the accretion rate, 
    {cannot be} 
    {excluded but} 
    but other accretion instabilities in the inner disk remain possible. Further, the disk warp may be produced by a time variable angular momentum vector of material accreted onto the disk, and could potentially drive the inner disk accretion instabilities. 
\end{itemize}
Regardless of whether the structure and variability of HOPS-358 are driven by gravitational instabilities, inner disk accretion instabilities, or unseen companions, the system certainly provides one of the best examples of how observations across a range of wavelengths (infrared to millimeter) and modes (unresolved flux monitoring to high resolution imaging) can help to illuminate the processes by which young stars grow.

\begin{acknowledgments}
{We thank the referee for a constructive report that helped to strengthen the contents of this work}. DJ is supported by NRC Canada and by an NSERC Discovery Grant.
CCP is supported by the National Research Foundation of Korea (NRF) grant funded by the Korean government (MEST) (No.\  2019R1A6A1A10073437).
SL, JEL, and CCP were supported by the NRF grant funded by the Korean government (MSIT) (grant numbers 2021R1A2C1011718 and RS-2024-00416859).
This paper makes use of the following ALMA data: ADS/JAO.ALMA\#2015.1.00041.S, ADS/JAO.ALMA\#2019.1.00691.S, 
ADS/JAO.ALMA\#2021.1.00844.S. ALMA is a
partnership of ESO (representing its member states), NSF (USA) and NINS (Japan), together with NRC
(Canada), MOST and ASIAA (Taiwan), and KASI (Republic of Korea), in cooperation with the Republic
of Chile. The Joint ALMA Observatory is operated by ESO, AUI/NRAO and NAOJ.
under cooperative agreement by Associated Universities, Inc.
This research has made use of the NASA/IPAC Infrared Science Archive, which is funded by the National Aeronautics and Space Administration and operated by the California Institute of Technology.
\end{acknowledgments}

\software{CASA \citep{TheCASATeam2022CASAAstronomy}, astropy \citep{Robitaille2013Astropy:Astronomy,Collaboration2018ThePackage,TheAstropyCollaboration2022ThePackage}, galario \citep{Tazzari2017GALARIO:Observations}, matplotlib \citep{Hunter2007Matplotlib:Environment}, dynesty \citep{Speagle2020Dynesty:Evidences}}

\bibliography{references,references_carlos}

\begin{deluxetable*}{lcclcc}
\tablecaption{JCMT Peak Brightness}
\tablenum{1}
\tabletypesize{\normalsize}
\label{table:jcmt_data}
\tablehead{\colhead{JD}        & \colhead{Flux450} & \colhead{Flux850} & \colhead{JD}        & \colhead{Flux450} & \colhead{Flux850} \\ \colhead{} & \colhead{(Jy beam$^{-1}$)} & \colhead{(Jy beam$^{-1}$)} & \colhead{} & \colhead{(Jy beam$^{-1}$)} & \colhead{(Jy beam$^{-1}$)}}
\startdata
57382.494 &         & 1.524   & 58915.295 & 4.981   & 1.186   \\
57403.347 & 7.118   & 1.538   & 59072.634 &         & 1.256   \\
57424.221 & 6.821   & 1.507   & 59087.644 & 5.261   & 1.27    \\
57447.188 & 6.541   & 1.477   & 59106.721 & 5.122   & 1.283   \\
57476.204 & 7.249   & 1.508   & 59120.762 & 4.929   & 1.29    \\
57505.216 & 6.219   & 1.52    & 59155.479 & 5.139   & 1.269   \\
57627.603 &         & 1.546   & 59180.37  & 5.192   & 1.259   \\
57712.623 &         & 1.559   & 59195.44  &         & 1.262   \\
57718.503 & 6.853   & 1.53    & 59212.543 &         & 1.284   \\
57790.195 &         & 1.426   & 59228.4   &         & 1.307   \\
57831.232 &         & 1.39    & 59242.188 &         & 1.289   \\
57864.22  &         & 1.429   & 59256.199 & 5.235   & 1.253   \\
57928.791 &         & 1.401   & 59275.273 & 5.362   & 1.262   \\
57965.695 &         & 1.345   & 59300.225 & 5.41    & 1.252   \\
57997.667 & 5.874   & 1.346   & 59321.217 & 5.566   & 1.272   \\
58025.62  & 5.358   & 1.312   & 59440.629 & 5.449   & 1.276   \\
58064.438 & 5.652   & 1.32    & 59454.633 & 5.697   & 1.26    \\
58095.422 & 4.846   & 1.294   & 59485.564 & 5.391   & 1.248   \\
58133.41  & 5.351   & 1.306   & 59511.48  & 6.053   & 1.325   \\
58166.197 & 4.535   & 1.259   & 59526.396 & 5.756   & 1.363   \\
58208.185 &         & 1.241   & 59590.233 &         & 1.431   \\
58367.606 &         & 1.266   & 59604.35  & 6.303   & 1.478   \\
58395.6   &         & 1.273   & 59623.331 & 6.086   & 1.453   \\
58423.639 &         & 1.344   & 59637.208 & 6.022   & 1.421   \\
58455.341 & 5.951   & 1.358   & 59652.223 & 5.999   & 1.466   \\
58486.269 & 5.131   & 1.244   & 59785.784 & 6.012   & 1.387   \\
58519.335 & 5.459   & 1.277   & 59818.693 & 6.214   & 1.434   \\
58552.3   &         & 1.265   & 59832.555 &         & 1.435   \\
58580.259 &         & 1.292   & 59852.547 & 6.036   & 1.464   \\
58715.623 &         & 1.159   & 59867.665 & 5.819   & 1.433   \\
58752.54  &         & 1.165   & 59884.462 &         & 1.411   \\
58774.726 & 4.88    & 1.173   & 59905.598 & 5.776   & 1.437   \\
58788.463 & 4.705   & 1.214   & 59936.383 & 6.083   & 1.405   \\
58836.357 & 4.643   & 1.149   & 59964.441 &         & 1.369   \\
58850.286 & 4.775   & 1.19    & 60001.195 & 6.029   & 1.425   \\
58872.39  & 4.386   & 1.131   & 60166.748 & 5.734   & 1.394   \\
58901.21  & 4.613   & 1.158   &           &         &        
\enddata
\end{deluxetable*}

\end{document}